%% file: main.tex
\documentclass[acmsmall,screen]{acmart}

\renewcommand\footnotetextcopyrightpermission[1]{} 
\usepackage[ruled,vlined,lined,commentsnumbered]{algorithm2e}
\usepackage{amsmath,amsfonts}
\usepackage{array}
\usepackage{booktabs}
\usepackage[skip=1pt,labelfont=bf]{caption}
 \usepackage{calligra}
\usepackage{color, colortbl}
\usepackage{courier}
\usepackage{csvsimple}
\usepackage{enumitem}
\usepackage{fancybox}
\usepackage{fontenc}
\usepackage{graphicx}
\usepackage{listings}
\usepackage{longtable}
\usepackage{lscape}
\usepackage{makecell}
\usepackage{marvosym}
\usepackage{moreverb}
\usepackage{multicol}
\usepackage{multirow}
\usepackage{pifont}
\usepackage{rotating}
\usepackage{setspace}
\usepackage{subfigure}
\usepackage[most]{tcolorbox}
\usepackage{threeparttable}
\usepackage{tikz}
\usepackage[normalem]{ulem}
\usepackage{url}
\usepackage{soul}
\usepackage{wasysym}
\usepackage{indentfirst}
\usepackage{textcomp}
\usepackage{xcolor}
\usepackage{wrapfig}
\usepackage{pdflscape}
\usepackage{hyperref}

\newcommand{\tabincell}[2]{\begin{tabular}{@{}#1@{}}#2\end{tabular}}
\usepackage{amsmath}
\usepackage{array}
\usepackage{balance}

\usepackage{microtype}

\newcolumntype{L}[1]{>{\raggedright\let\newline\\\arraybackslash\hspace{0pt}}m{#1}}
\newcolumntype{C}[1]{>{\centering\let\newline\\\arraybackslash\hspace{0pt}}m{#1}}
\newcolumntype{R}[1]{>{\raggedleft\let\newline\\\arraybackslash\hspace{0pt}}m{#1}}

\newboolean{showcomments}
\setboolean{showcomments}{true}
\ifthenelse{\boolean{showcomments}}
 { \newcommand{\mynote}[2]{
      \fbox{\bfseries\sffamily\scriptsize#1}
        {\small$\blacktriangleright$\textsf{\emph{#2}}$\blacktriangleleft$}}}
        { \newcommand{\mynote}[2]{}}

\definecolor{darkgreen}{rgb}{114,169,119}

\newcommand{\intuition}[1]{
\begin{tcolorbox}[colback=white,boxrule=1pt,top=0pt,bottom=0pt,left=1pt,right=2pt,top=2pt,bottom=2pt]
\em #1
\end{tcolorbox}
}

\newcommand{\find}[1]{
\begin{tcolorbox}[leftrule=1mm,toprule=0mm,bottomrule=0mm,left=1pt,right=2pt,top=2pt,bottom=2pt]
\em #1
\end{tcolorbox}
}

\definecolor{lightgray}{gray}{0.9}

\AtBeginDocument{%
  \providecommand\BibTeX{{%
    \normalfont B\kern-0.5em{\scshape i\kern-0.25em b}\kern-0.8em\TeX}}}

\setcopyright{none}

\begin{document}

\title{Pros and Cons! Evaluating ChatGPT on Software Vulnerability
}

\author{Xin Yin}
\affiliation{
  \institution{Zhejiang University}
  \city{Hangzhou}
  \state{Zhejiang}
  \country{China}}
\email{xyin@zju.edu.cn}

\begin{abstract}

This paper proposes a pipeline for quantitatively
evaluating interactive LLMs such as ChatGPT using publicly available dataset. 
We carry out an extensive technical evaluation of ChatGPT using Big-Vul covering five different common software vulnerability tasks. 
We evaluate the multitask and multilingual aspects of ChatGPT based on this dataset.
We found that the existing state-of-the-art methods are generally superior to ChatGPT in software vulnerability detection. 
Although ChatGPT improves accuracy when providing context information, it still has limitations in accurately predicting severity ratings for certain CWE types. 
In addition, ChatGPT demonstrates some ability in locating vulnerabilities for certain CWE types, but its performance varies among different CWE types. 
ChatGPT exhibits limited vulnerability repair capabilities in both providing and not providing context information.
Finally, ChatGPT shows uneven performance in generating CVE descriptions for various CWE types, with limited accuracy in detailed information.
Overall, though ChatGPT performs well in some aspects, it still needs improvement in understanding the subtle differences in code vulnerabilities and the ability to describe vulnerabilities in order to fully realize its potential.
Our evaluation framework provides valuable insights for further enhancing ChatGPT's software vulnerability handling capabilities.

\end{abstract}

\maketitle

\input{sections/introduction}
\label{introduction}

\input{sections/background_relatedwork}
\label{background_relatedwork}

\input{sections/experimental_setup}
\label{experimental_setup}

\input{sections/evaluation_results}

\label{evalution_results}

\input{sections/threats_to_validation}
\label{threats_to_validation}

\input{sections/conclusion}
\label{conclusion}

\balance
\bibliographystyle{ACM-Reference-Format}
\bibliography{main}

\end{document}

%% file: sections/introduction.tex
\section{Introduction}

\begin{figure}[htbp]
    \centering
    \includegraphics[width=1\linewidth]{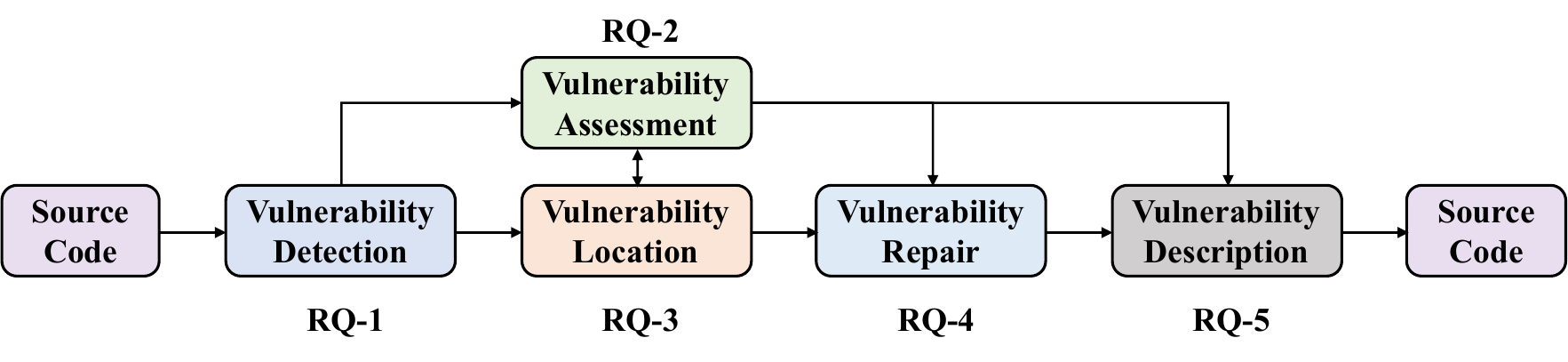}
    \caption{The relationship among software vulnerability activities.}
    \label{fig:relationship}
\end{figure}

Software Vulnerabilities (SVs) can expose software systems to risk situations and eventually causes huge economic losses or even threatens people's lives.
Therefore, completing software vulnerabilities is an important task for software quality assurance (SQA).
Generally, there are many important software quality activities for software vulnerabilities such as SV detection, SV assessment, SV location, SV repair, and SV description generation.
The relationship among the SQA activities can be illustrated in Fig.~\ref{fig:relationship}.

Meanwhile, ChatGPT has become a well-known AI model for general public users including users who are not familiar with LLMs or computer science technology.
It is developed by OpenAI and technically built on GPT-3.5 architecture~\cite{bang2023multitask}.
The most important reason why ChatGPT has become so popular is that ChatGPT has the ability to cover innumerable use cases for both academic and non-academic users online.
Besides, some discussions are conducted about whether ChatGPT is approaching Artificial General Intelligence (AGI) since it has the ability to address many types of tasks without specific fine-tune~\cite{remarkonllm,firendorfoe}. 
There still have some debates concerning its failures in a few simple tasks~\cite{choi2023chatgpt,gilson2022well,shen2023chatgpt}.

In this paper, we focus on evaluating ChatGPT's performance in various software vulnerability (SV)-related tasks in a zero-shot setting to obtain a basic, comprehensive, and better understanding of its multi-task ability, and we aim to answer the following research questions.

\begin{itemize}[leftmargin=*]
    \item \textbf{RQ-1: How does ChatGPT perform on vulnerability detection?}  Software Vulnerabilities (SVs) can expose software systems to risk situations and consequently software function failure. Therefore, detecting these SVs is an important task for software quality assurance. We want to explore the ability of ChatGPT on vulnerability detection as well as the performance difference compared with state-of-the-art approaches.

    \item \textbf{RQ-2: How does ChatGPT perform on vulnerability assessment?} In practice, due to the limitation of SQA resources~\cite{khan2018review}, it is impossible to treat all detected SVs equally and fix all SVs simultaneously. Thus, it is necessary to prioritize these detected software vulnerabilities for better treatment. An effective solution to prioritize those SVs, which has imminent and serious threats to the systems of interest, is to use one of the most widely known SV assessment framework CVSS (Common Vulnerability Scoring System)~\cite{le2021survey}, which characterizes SVs by considering three metric groups: Base, Temporal and Environmental. The metrics that are in the groups can be further used as the criterion for selecting serious SVs to fix early. Therefore, we want to explore the ability of ChatGPT on  assessing vulnerabilities.

    \item \textbf{RQ-3: How does ChatGPT perform on vulnerability location?} Identifying the precise location of vulnerabilities in software systems is of critical importance for mitigating risks and improving software quality. The vulnerability location task involves  pinpointing these weaknesses accurately and helps to narrow the scope for developers to fix problems. Therefore, we aim to investigate ChatGPT's capability in effectively identifying the precise location of vulnerabilities in software systems.

    \item \textbf{RQ-4: How does ChatGPT perform on vulnerability repair?}
    Addressing vulnerabilities in software systems is crucial for risk mitigation and software quality enhancement. Vulnerability repair task requires effective identification and remediation of these flaws with source code and consequently helps to improve the development efficiency.
    We want to examine ChatGPT's proficiency in efficiently repairing vulnerabilities in software systems.

    \item \textbf{RQ-5: How does ChatGPT perform on vulnerability description generation?}
    Understanding the intricacies of vulnerabilities in software systems plays a pivotal role in alleviating risks and bolstering software quality. The vulnerability description task focuses on conveying a detailed explanation of these identified issues in the source codes and helps participants to better understand the risk as well as its impacts. Our aim is to assess ChatGPT's capacity to effectively generate the description of vulnerabilities within software systems.

\end{itemize}

Besides, to extensively and comprehensively analyze the model's ability, we use large-scale dataset containing real-world project vulnerabilities (named Big-Vul).
Then, we carefully design experiments to discover the findings by answering five RQs.
Eventually, the main contribution of our work is summarized as follows and takeaway findings are shown in Table~\ref{tab:takeaway}.
\begin{itemize}[leftmargin=*]
    \item We extensively evaluate the performance of ChatGPT on different software vulnerability tasks and conduct an extensive comparison among ChatGPT and learning-based approaches on software vulnerability.
    \item We design five RQs to comprehensively understand ChatGPT from different dimensions, and supported our conclusions with effective results and examples.
    \item We release our reproduction package for further study~\cite{replication}.
\end{itemize}

\begin{table*}[!th]\vspace{-0.3cm}
    \centering
    \caption{Insights and Takeaways: Evaluating ChatGPT on Software Vulnerability}
    
    \resizebox{1.0\linewidth}{!}
    {
      \begin{tabular}{l|l}
      \toprule
      \textbf{Dimension} &{\textbf{Findings or Insights}}\\
      \midrule
      \textbf{Vulnerability Detection} & 

      \tabincell{l}{
          \textbf{1}. The existing state-of-the-art approaches perform better than ChatGPT. \\
          \textbf{2}. ChatGPT excels at null pointer and access control vulnerabilities but struggles with others. \\
          \textbf{3}. ChatGPT is easily swayed to change vulnerability classifications, indicating low confidence.
      } \\
\midrule
       \textbf{Vulnerability Assessment} & \tabincell{l}{
       \textbf{4}. ChatGPT has the limited capacity for assessment of vulnerability, but can be improved with more context.\\
       \textbf{5}. ChatGPT's inconsistent vulnerability severity prediction across CWEs highlights needed enhancements. \\
       \textbf{6}. ChatGPT's vulnerability severity prediction improves with key info, but varies by CWE type.} \\
       \midrule
      \textbf{Vulnerability Localization}& \tabincell{l}{
      \textbf{7}. ChatGPT exhibits a certain capability in vulnerability locations and its performance varies among different CWE types.
      } \\
      \midrule
      \textbf{Vulnerability Repair}& \tabincell{l}{
      \textbf{8}. ChatGPT has limited ability in repairing vulnerability no matter when provided with context information or not.\\
      \textbf{9}. ChatGPT’s performance in repairing can be attributed to its ability to recognize and understand specific vulnerability patterns. 
       }\\
       \midrule
        \textbf{Vulnerability Description}& \tabincell{l}{
        \textbf{10}. ChatGPT has uneven CVE description generation across CWEs, with limited detail accuracy. \\
        \textbf{11}. ChatGPT has uneven vulnerability description ability due to training data, complexity, and language specificity.
        }  \\
         \bottomrule
       \end{tabular}
       }\vspace{-0.3cm}
       \label{tab:takeaway}
\end{table*}

%% file: sections/background_relatedwork.tex
\section{Background and Related Work}

\subsection{Large Language Model}
{
Since the advancements in Natural Language Processing, Large Language Models (LLMs)~\cite{brown2020language} have seen widespread adoption due to their capacity to be effectively trained with billions of parameters and training samples, resulting in significant performance enhancements.
LLMs can readily be applied to downstream tasks through either fine-tuning~\cite{radford2018improving} or prompting~\cite{liu2023pre}.
Their versatility stems from being trained to possess a broad understanding, enabling them to capture diverse knowledge across various domains.
Fine-tuning involves updating the model parameters specifically for a given downstream task through iterative training on a specific dataset. 
In contrast, prompting allows for direct utilization by providing natural language descriptions or a few examples of the downstream task.
Compared to prompting, fine-tuning is resource-intensive as it necessitates additional model training and is applicable in limited scenarios, particularly when adequate training datasets are unavailable.
}

LLMs are usually built on the transformer architecture~\cite{vaswani2017attention} and can be classified into three types of architectures: encoder-only, encoder-decoder, and decoder-only. 
Encoder-only (e.g., CodeBERT~\cite{feng2020codebert}, GraphCodeBERT~\cite{guo2020graphcodebert}, and UniXcoder~\cite{guo2022unixcoder}) and Encoder-Decoder (e.g., PLBART~\cite{ahmad2021unified} and CodeT5~\cite{wang2021codet5}) models are trained using Masked Language
Modeling (MLM) or Masked Span Prediction (MSP) objective, respectively, where a small portion (e.g., 15\%) of the tokens are replaced with either masked tokens or masked span tokens
LLMs are trained to recover the masked tokens.
These models are trained as general ones on the code-related data and then are fine-tuned for the downstream tasks to achieve superior performance. 
Decoder-only models also attract a small portion of people's attention and they are trained by using Causal Language Modeling objectives to predict the probability of the next token given all previous tokens. 
GPT~\cite{radford2018improving} and its variants are the most representative models, which bring the large language models into practical usage.

Recently, the ChatGPT model attracts the widest attention from the world, which is the successor of the large language model InstructGPT~\cite{ouyang2022training} with a dialog interface that is fine-tuned using the Reinforcement Learning with Human Feedback (RLHF) approach~\cite{christiano2017deep,ouyang2022training,ziegler2019fine}.
RLHF initially fine-tunes the base model using a small dataset of prompts as input and the desired output, typically human-written, to refine its performance.
Subsequently, a reward model is trained on a larger set of prompts by sampling outputs generated by the fine-tuned model. 
These outputs are then reordered by human labelers to provide feedback for training the reward model.
Reinforcement learning~\cite{schulman2017proximal} is then used to calculate rewards for each output generated based on the reward model, updating LLM parameters accordingly. 
With fine-tuning and alignment with human preferences, LLMs better understand input prompts and instructions, enhancing performance across various tasks~\cite{bang2023multitask,ouyang2022training}.

\subsection{Software Vulnerability}
{
Software Vulnerabilities (SVs) can expose software systems to risk situations and consequently make the software under cyber-attacks, eventually causing huge economic losses and even threatening people's lives.
Therefore, vulnerability databases have been created to document and analyze publicly known security vulnerabilities.
For example, Common Vulnerabilities and Exposures (CVE)~\cite{cve,bhandari2021cvefixes} and SecurityFocus~\cite{securityfocus} are two well-known vulnerability databases.
Besides, Common Weakness Enumeration (CWE) defines the 
common software weaknesses of individual vulnerabilities, which are often referred to as vulnerability types of CVEs. 
To better address these vulnerabilities, researchers have proposed many approaches for understanding the effects of software vulnerabilities, including SV detection~\cite{zhou2019devign,fu2022linevul,cao2022mvd,li2021sysevr,cheng2022path,wu2022vulcnn,li2018vuldeepecker,hin2022linevd,zhan2021atvhunter,chakraborty2021deep,ni2023distinguishing}, SV assessment~\cite{ni2023fva,feutrill2018effect,le2021survey,spanos2018multi,le2021deepcva}, SV localization~\cite{li2021vuldeelocator,li2021vulnerability,ni2022best}, SV repair~\cite{ni2022defect,zhang2022program,chen2019sequencer,zhu2021syntax} as well as SV description~\cite{sun2021generating,guo2022detecting,guo2021key,guo2020predicting}.
Many novel technologies are adopted to promote the progress of software vulnerability management, including software analysis~\cite{fan2019smoke,li2020pca}, machine learning~\cite{zhou2019devign,hin2022linevd}, and deep learning~\cite{ni2023fva,li2021vulnerability}, especially  large language models~\cite{guo2022detecting,guo2021key,yin2024multitask}.
}

%% file: sections/experimental_setup.tex
\section{Experimental Design}

In this section, we present our research question (RQs), our studied dataset, the baseline approaches, and the evaluation metrics.

\subsection{Studied Dataset}
\label{lab:dataset}

We adopt the widely used dataset (named Big-Vul) provided by Fan et al.~\cite{fan2020ac} by considering the following reasons. 
The most important one is to satisfy the distinct characteristics of the real world as well as the diversity in the dataset, which is suggested by previous works~\cite{hin2022linevd,chakraborty2021deep}.
Big-Vul, to the best of our knowledge,  is the most large-scale vulnerability dataset with diverse information about the vulnerabilities, which are collected from practical projects and these vulnerabilities are recorded in the Common Vulnerabilities and Exposures (CVE)\footnote{https://cve.mitre.org/}.
The second one is to compare fairly with existing state-of-the-art (SOTA) approaches (e.g., LineVul, Devign).
The third one is Big-Vul is the only one vulnerability dataset that contains the fixed version of vulnerable functions, which can be utilized to 
evaluate whether ChatGPT has the ability to fix a vulnerable function.

Big-Vul totally contains 3,754 code vulnerabilities collected from 348 open-source projects spanning 91 different vulnerability types from 2002 to 2019.
It has 188,636 C/C++ functions with a vulnerable ratio of 5.7\% (i.e., 10,900 vulnerability functions).
The authors linked the code changes with CVEs and their descriptive information to enable a deeper analysis of the vulnerabilities.

The original dataset cannot be directly used in our study. 
Therefore, we filter the original dataset from two aspects.
Firstly, considering the input length limitation (i.e., 4,096 tokens) of ChatGPT API, we use the \textit{tiktoken tool}\footnote{https://github.com/openai/tiktoken}, a fast BPE tokenizer for use with OpenAI's models, to filter out functions that exceed 2,000 tokens and leave the remaining 2,096 tokens as the input \textit{prompt} as well as the output of fixed functions of a vulnerable function generated by ChatGPT.
Secondly, we filter out the instances whose CWE ID is recorded as NaN.

Finally, to better evaluate ChatGPT's ability and considering the large consumption of interaction with ChatGPT,
we need to determine the number of instances (i.e., functions in Big-Vul) in order to get results (i.e., testing dataset) that reflect the target dataset (i.e., Big-Vul) as precisely as needed. 
We sample the testing instances for each CWE in Big-Vul with 95\% confidence and 5\% interval\footnote{https://surveysystem.com/sscalc.htm}.
Eventually, we obtain 724 vulnerable functions and 12,701 non-vulnerable functions to conduct our study.
The statistics are shown in Table~\ref{tab:dataset}.

\begin{table}[htbp]
  \centering
  \caption{The statistic of studied dataset}
  {
    \begin{tabular}{|l|r|r|c|}
    \hline
    \textbf{Dataset} & \textbf{\# Vulnerablities} & \textbf{\# Non-Vulnerablities} & \textbf{ \% Ratio} \\
    \hline
    Original Big-Vul & 10,900  & 177,736  & 1:16.3 \\
    \hline
    Filtered Big-Vul  & 8,048  & 142,302  & 1:17.9 \\
    \hline
   \rowcolor{lightgray} Sampled Big-Vul & 724   & 12,701  & 1:17.5 \\
    \hline
    \end{tabular}%
    }
  \label{tab:dataset}%
\end{table}%

\subsection{Evaluation Metrics}

For considered software vulnerability-related tasks, we will perform evaluations using the
widely adopted performance metrics.
For software vulnerability detection tasks, metrics such as Precision, Recall, F1-score, and Accuracy are used~\cite{ni2022best}.
For the software vulnerability assessment, Accuracy is adopted.
For the software vulnerability location task, we adopt three performance metrics: Hit@Accuracy, Precision, and Recall.
For the software vulnerability repair task, we use Hit@Accuracy.
For the software vulnerability description generation task, we use Rouge-1, Rouge-2, and Rouge-L metrics.

\subsection{Baselines}

To comprehensively compare the performance of ChatGPT with existing state-of-the-art (SOTA) approach, in this study, we consider the four approaches: Devign~\cite{zhou2019devign}, {\sc ReVeal}~\cite{chakraborty2021deep}, {\sc IVDetect}~\cite{li2021vulnerability}, and LineVul~\cite{fu2022linevul}.
We briefly introduce them as follows.

\textbf{Devign} proposed by Zhou et al.~\cite{zhou2019devign} is a general graph neural network-based model for graph-level classification through learning on a rich set of code semantic representations including AST, CFG, DFG, and code sequences.
It uses a novel $Conv$ module to efficiently extract useful features in the learned rich node representations for graph-level classification.

{\sc \textbf{ReVeal}} proposed by Chakraborty et al.~\cite{chakraborty2021deep} contains two main phases. 
In the feature extraction phase, it translates code into a graph embedding, and in the training phase, it trains a representation learner on the extracted features to obtain a model that can distinguish the vulnerable functions from non-vulnerable ones.

{\sc \textbf{IVDetect}} proposed by Li et  al.~\cite{li2021vulnerability} contains the coarse-grained vulnerability detection component and fine-grained interpretation component.
In particular, {\sc IVDetect} represents source code in the form of a program dependence graph (PDG) and treats the vulnerability detection problem as graph-based classification via graph convolution network with feature attention.
As for interpretation, {\sc IVDetect} adopts a GNNExplainer to provide fine-grained interpretations that include the sub-graph in PDG with crucial statements that are relevant to the detected vulnerability.

\textbf{LineVul} proposed by Fu et al.~\cite{fu2022linevul} is a Transformer-based line-level vulnerability prediction approach. 
LineVul leverages BERT architecture with self-attention layers which can capture long-term dependencies within a long sequence. 
Besides, benefiting from the large-scale pre-trained model, LineVul can intrinsically capture more lexical and logical semantics for the given code input.
Moreover, LineVul adopts the attention mechanism of BERT architecture to locate the vulnerable lines for finer-grained detection.

\subsection{Implementation}
\label{sec:implementation}

We implemented experiment in Python by wrapping the state-of-the-art ChatGPT~\cite{openai2022chatgpt} model through its API support~\cite{2023chatgptendpoint} and adhere to the best-practice guide~\cite{shieh2023best} for each prompt.
We utilize the gpt-3.5-turbo-0301 model from the ChatGPT family, which is the version used uniformly for our experiments.
We also instruct ChatGPT to output the results in JSON format to facilitate the automatic organization of the data.
Regarding ReVeal, IVDetect, Devign and LineVul, we utilize their publicly available source code and perform fine-tuning with the default parameters provided in their original code. 
Considering Devign's code is not publicly available, we make every effort to replicate its functionality and achieve similar results on the original paper's dataset.
All these models are implemented using the PyTorch~\cite{pytorch} framework.
The evaluation is conducted on a 16-core workstation equipped with an Intel(R) Xeon(R) Gold 6226R CPU @ 2.90Ghz, 192GB RAM, running Ubuntu 20.04.1 LTS.

%% file: sections/evaluation_results.tex
\section{Experimental Results}
This section presents the experimental results of ChatGPT by evaluating ChatGPT performances on the widely used comprehensive  dataset (i.e., Big-Vul~\cite{fan2020ac} covering five SV-related tasks.

\begin{table}[htbp]
  \centering
  \caption{The comparison between ChatGPT and four approaches on software vulnerability detection}
  {
  \begin{threeparttable}

    \begin{tabular}{|l|l|r|c|c|c|c|}
    \hline
    \textbf{Setting} & \textbf{Approaches} & \textbf{\# Testing} & \textbf{F1-score} & \textbf{Precision} & \textbf{Recall} & \textbf{Accuracy} \\
    \hline \hline
    \multirow{5}[2]{*}{\textbf{Sampled}} & {ChatGPT} &             13,425  & 0.106 & 0.091 & 0.127 & 0.884 \\
\cline{2-7}          & {LineVul} &             13,425  & \textbf{0.260} & \textbf{0.340} & 0.210 & \textbf{0.935} \\
\cline{2-7}          & {Devign} &               9,115  & 0.162 & 0.094 & {0.571} & 0.674 \\
\cline{2-7}          & {ReVeal} &               9,115  & 0.176 & 0.108 & 0.487 & 0.749 \\
\cline{2-7}          & {IVDetect} &               9,195  & 0.181 & 0.110 & 0.498 & 0.741 \\
    \hline \hline

\multirow{8}[4]{*}{\textbf{ GPT\&EB} } & {ChatGPT} &    \multirow{2}[1]{*}{13,425}  & 0.106 & 0.091 & 0.127 & 0.884 \\
        & {LineVul} &          & \textbf{0.260} & \textbf{0.340} & 0.210 & \textbf{0.935} \\

 \cline{2-7} & {ChatGPT} &   \multirow{2}[1]{*}{9,115}  & 0.109 & 0.091 & 0.135 & \textbf{0.878} \\
        & {Devign} &           & \textbf{0.162} & \textbf{0.094} & 0.571 & 0.674 \\
\cline{2-7}          & {ChatGPT} &   \multirow{2}[1]{*}{9,115}  & 0.109 & 0.091 & 0.135 & \textbf{0.878} \\
         & {ReVeal} &       & \textbf{0.176} & \textbf{0.108} & \textbf{0.487} & 0.749 \\
\cline{2-7}          & {ChatGPT} &  \multirow{2}[1]{*}{ 9,195}  & 0.113 & 0.095 & 0.138 & \textbf{0.875} \\
         & {IVDetect} &       & \textbf{0.181} & \textbf{0.110} & \textbf{0.498} & 0.741 \\
    \hline \hline
    \multirow{5}[1]{*}{
    \textbf{GPT\&AB}}
     & {ChatGPT} & \multirow{5}[2]{*}{9,025}  & 0.108 & 0.090 & 0.133 & 0.877 \\
\cline{4-7}          & {LineVul} &                & \textbf{0.240} & \textbf{0.350} & 0.180 & \textbf{0.940} \\
\cline{4-7}          & {Devign} &         & 0.163 & 0.095 & \textbf{0.572} & 0.673 \\
\cline{4-7}          & {ReVeal} &        & 0.176 & 0.108 & 0.488 & 0.746 \\
\cline{4-7}          & {IVDetect} &        & 0.175 & 0.107 & 0.484 & 0.745 \\
    \hline
    \end{tabular}%
    ``GPT\&EB'' refers to ChatGPT $\cap$ Each Baseline.\\
    ``GPT\&AB'' refers to ChatGPT $\cap$ All Baselines.
      \end{threeparttable}
    }
  \label{tab:rq1-overall}%
\end{table}%

\subsection{\bf RQ-1: Evaluating Vulnerability Detection of ChatGPT}
In this RQ, we first investigate the vulnerability detection of ChatGPT and make a comparison with the existing state-of-the-art (SOTA) approaches.
Then, we want to investigate whether the decision results of ChatGPT can be induced by human instructions.

\noindent
{\bf {\em  RQ-1.1: How does ChatGPT perform on vulnerability detection compared to the state-of-the-art approaches?}}

\noindent
\textbf{Experimental Setting}.
We sample each CWE from the Big-Vul dataset as introduced in Section~\ref{lab:dataset} and adopt the widely used four performance measures (i.e., Precision, Recall, F1-score, and Accuracy) for evaluating the performance of ChatGPT on detecting vulnerability.
Notice that, we use the following instruction for this SV detection task.

\noindent
\intuition{
\em{\textbf{Prompt 1.1}}: I want you to act as a vulnerability scanner. I will provide you a C code snippet and want you to tell whether it has a vulnerability. You need to output ``yes'' or ``no'' first (output no if uncertain), and then explain.
}

Moreover, to better understand the performance difference between ChatGPT and SOTA software vulnerability detection approaches, we totally consider four approaches which can be divided  into two groups: transformer-based (LineVul) and graph-based (Devign, ReVeal, IVdetect). 
Besides, graph-based approaches need to obtain the structure information (e.g., control flow graph (CFG), data flow graph (DFG)) of the studied functions, so we adopt the same toolkit with \textit{Joern}~\cite{joern} to transform sampled functions.
The functions are dropped out directly if they cannot be transformed by \textit{Joern} successfully.
Finally, we fine-tune the four SOTAs on the remaining dataset of Big-Vul after sampling the instances for testing.
That is, we refer to Big-Vul as $D$ and the sampled dataset from $D$ is referred to as $S$, therefore the difference between $D$ and $S$ is used to fine-tune the four approaches.

Besides, considering the difference in the testing size of each approach, we show the three types of settings when comparing ChatGPT and baselines: (1) \textit{Sampled} means  evaluating each approach on the sampled functions that can be successfully analyzed; 
(2) \textit{Intersection between ChatGPT and each baseline (ChatGPT $\cap$ 
 Each Baseline)} means evaluating the performance difference between ChatGPT and each baseline on the intersection of the two approaches' testing dataset; (3) \textit{Intersection among ChatGPT and all baselines (ChatGPT $\cap$ All Baselines)} means evaluating the performance difference  among ChatGPT and all baseline on the intersection of all approaches' testing dataset.

{
Apart from presenting the overall performance comparison, we also give the detailed performance of ChatGPT on the Top-10 CWE types in the testing dataset for a better analysis as shown in Table.~\ref{tab:chatgpt-vd-top-10}.
}

\begin{table*}[htbp]
  \centering
   \caption{The software vulnerability detection performance comparison between ChatGPT and four baselines}
  \resizebox{\linewidth}{!}{
    \begin{tabular}{|lcc||c|cccc||c|cccc|}
    \hline
    \multirow{2}[1]{*}{\textbf{CWE Type}} & \multirow{2}[1]{*}{\textbf{\# Sampled}} & \multirow{2}[1]{*}{\textbf{\# Vulnerability}} & \textbf{ChatGPT} & \multicolumn{1}{c|}{\textbf{LineVul}} & \multicolumn{1}{c|}{\textbf{Devign}} & \multicolumn{1}{c|}{\textbf{ReVeal}} & \textbf{IVDetect} & \textbf{ChatGPT} & \multicolumn{1}{c|}{\textbf{LineVul}} & \multicolumn{1}{c|}{\textbf{Devign}} & \multicolumn{1}{c|}{\textbf{ReVeal}} & \textbf{IVDetect} \\
\cline{4-13}          &       &       & \multicolumn{5}{c||}{\textbf{F1-score}} & \multicolumn{5}{c|}{\textbf{Precision}} \\
    \hline
    CWE-125 & 305   & 16    & 0.06  & 0.23  & 0.17  & 0.19  & \cellcolor[rgb]{ .851,  .882,  .949}\textbf{0.29 } & 0.05  & \cellcolor[rgb]{ .851,  .882,  .949}\textbf{0.30 } & 0.10  & 0.11  & 0.18  \\
    CWE-476 & 302   & 13    & \cellcolor[rgb]{ .988,  .894,  .839}\textbf{0.19 } & 0.10  & 0.10  & 0.14  & 0.08  & \cellcolor[rgb]{ .988,  .894,  .839}\textbf{0.17 } & 0.14  & 0.06  & 0.08  & 0.05  \\
    CWE-189 & 248   & 11    & 0.06  & 0.21  & 0.11  & 0.18  & \cellcolor[rgb]{ .851,  .882,  .949}\textbf{0.25 } & 0.04  & \cellcolor[rgb]{ .851,  .882,  .949}\textbf{0.25 } & 0.06  & 0.11  & 0.15  \\
    CWE-200 & 231   & 10    & 0.00  & 0.13  & 0.17  & 0.14  & \cellcolor[rgb]{ .851,  .882,  .949}\textbf{0.18 } & 0.00  & \cellcolor[rgb]{ .851,  .882,  .949}\textbf{0.20 } & 0.10  & 0.09  & 0.11  \\
    CWE-119 & 221   & 9     & 0.08  & \cellcolor[rgb]{ .851,  .882,  .949}\textbf{0.43 } & 0.15  & 0.18  & 0.25  & 0.07  & 0.60  & 0.08  & 0.10  & \cellcolor[rgb]{ .851,  .882,  .949}\textbf{0.15 } \\
    CWE-264 & 216   & 9     & \cellcolor[rgb]{ .988,  .894,  .839}\textbf{0.16 } & 0.00  & 0.03  & 0.11  & 0.10  & \cellcolor[rgb]{ .988,  .894,  .839}\textbf{0.12 } & 0.00  & 0.02  & 0.06  & 0.06  \\
    CWE-362 & 212   & 10    & 0.21  & 0.00  & 0.13  & \cellcolor[rgb]{ .851,  .882,  .949}\textbf{0.27 } & \cellcolor[rgb]{ .851,  .882,  .949}\textbf{0.27 } & \cellcolor[rgb]{ .988,  .894,  .839}\textbf{0.17 } & 0.00  & 0.08  & \cellcolor[rgb]{ .851,  .882,  .949}\textbf{0.17 } & \cellcolor[rgb]{ .851,  .882,  .949}\textbf{0.17 } \\
    CWE-20 & 172   & 7     & 0.00  & \cellcolor[rgb]{ .851,  .882,  .949}\textbf{0.31 } & 0.10  & 0.18  & 0.12  & 0.00  & \cellcolor[rgb]{ .851,  .882,  .949}\textbf{0.33 } & 0.06  & 0.12  & 0.07  \\
    CWE-399 & 168   & 11    & 0.00  & 0.14  & \cellcolor[rgb]{ .851,  .882,  .949}\textbf{0.19 } & 0.13  & 0.14  & 0.00  & \cellcolor[rgb]{ .851,  .882,  .949}\textbf{0.33 } & 0.12  & 0.09  & 0.09  \\
    CWE-416 & 162   & 10    & 0.00  & 0.00  & \cellcolor[rgb]{ .851,  .882,  .949}\textbf{0.18 } & 0.15  & 0.17  & 0.00  & 0.00  & \cellcolor[rgb]{ .851,  .882,  .949}\textbf{0.12 } & 0.10  & \cellcolor[rgb]{ .851,  .882,  .949}\textbf{0.12 } \\
    
    \hline \hline
    \multirow{2}[1]{*}{\textbf{CWE Type}} & \multirow{2}[1]{*}{\textbf{\# Sampled}} & \multirow{2}[1]{*}{\textbf{\# Vulnerability}} & \textbf{ChatGPT} & \multicolumn{1}{c|}{\textbf{LineVul}} & \multicolumn{1}{c|}{\textbf{Devign}} & \multicolumn{1}{c|}{\textbf{ReVeal}} & \textbf{IVDetect} & \textbf{ChatGPT} & \multicolumn{1}{c|}{\textbf{LineVul}} & \multicolumn{1}{c|}{\textbf{Devign}} & \multicolumn{1}{c|}{\textbf{ReVeal}} & \textbf{IVDetect} \\
\cline{4-13}          &       &       & \multicolumn{5}{c||}{\textbf{Recall}}  & \multicolumn{5}{c|}{\textbf{Accuracy}} \\
    \hline
    CWE-125 & 305   & 16    & 0.06  & 0.19  & 0.56  & 0.69  & \cellcolor[rgb]{ .851,  .882,  .949}\textbf{0.88 } & 0.89  & \cellcolor[rgb]{ .851,  .882,  .949}\textbf{0.93 } & 0.70  & 0.69  & 0.78  \\
    CWE-476 & 302   & 13    & 0.23  & 0.08  & 0.38  & \cellcolor[rgb]{ .851,  .882,  .949}\textbf{0.54 } & 0.23  & 0.92  & \cellcolor[rgb]{ .851,  .882,  .949}\textbf{0.94 } & 0.70  & 0.72  & 0.78  \\
    CWE-189 & 248   & 11    & 0.09  & 0.18  & 0.45  & 0.55  & \cellcolor[rgb]{ .851,  .882,  .949}\textbf{0.73 } & 0.87  & \cellcolor[rgb]{ .851,  .882,  .949}\textbf{0.94 } & 0.67  & 0.78  & 0.80  \\
    CWE-200 & 231   & 10    & 0.00  & 0.10  & \cellcolor[rgb]{ .851,  .882,  .949}\textbf{0.70 } & 0.40  & 0.50  & 0.88  & \cellcolor[rgb]{ .851,  .882,  .949}\textbf{0.94 } & 0.70  & 0.79  & 0.81  \\
    CWE-119 & 221   & 9     & 0.11  & 0.33  & 0.67  & 0.67  & \cellcolor[rgb]{ .851,  .882,  .949}\textbf{0.78 } & 0.90  & \cellcolor[rgb]{ .851,  .882,  .949}\textbf{0.96 } & 0.68  & 0.75  & 0.81  \\
    CWE-264 & 216   & 9     & 0.22  & 0.00  & 0.11  & \cellcolor[rgb]{ .851,  .882,  .949}\textbf{0.33 } & \cellcolor[rgb]{ .851,  .882,  .949}\textbf{0.33 } & 0.90  & \cellcolor[rgb]{ .851,  .882,  .949}\textbf{0.94 } & 0.74  & 0.77  & 0.76  \\
    CWE-362 & 212   & 10    & 0.30  & 0.00  & 0.40  & \cellcolor[rgb]{ .851,  .882,  .949}\textbf{0.70 } & \cellcolor[rgb]{ .851,  .882,  .949}\textbf{0.70 } & 0.90  & \cellcolor[rgb]{ .851,  .882,  .949}\textbf{0.94 } & 0.75  & 0.82  & 0.82  \\
    CWE-20 & 172   & 7     & 0.00  & 0.29  & \cellcolor[rgb]{ .851,  .882,  .949}\textbf{0.43 } & \cellcolor[rgb]{ .851,  .882,  .949}\textbf{0.43 } & \cellcolor[rgb]{ .851,  .882,  .949}\textbf{0.43 } & 0.90  & \cellcolor[rgb]{ .851,  .882,  .949}\textbf{0.95 } & 0.70  & 0.84  & 0.73  \\
    CWE-399 & 168   & 11    & 0.00  & 0.09  & \cellcolor[rgb]{ .851,  .882,  .949}\textbf{0.45 } & 0.27  & 0.27  & 0.88  & \cellcolor[rgb]{ .851,  .882,  .949}\textbf{0.93 } & 0.74  & 0.76  & 0.77  \\
    CWE-416 & 162   & 10    & 0.00  & 0.00  & \cellcolor[rgb]{ .851,  .882,  .949}\textbf{0.40 } & 0.30  & 0.30  & 0.86  & \cellcolor[rgb]{ .851,  .882,  .949}\textbf{0.93 } & 0.78  & 0.79  & 0.82  \\
    \hline
    \end{tabular}%
    }
  \label{tab:chatgpt-vd-top-10}%
\end{table*}%

\textbf{Results}.
Table~\ref{tab:rq1-overall} shows the overall performance measures between ChatGPT and four approaches and the best performances are highlighted in bold.
The second column lists the number of testing sizes for each approach after pre-processing operation on the function.
Notice that, since both ChatGPT and LineVul have no requirement for a given function, they have the largest testing size.

{
As shown in Table~\ref{tab:rq1-overall}, we can obtain the following observations:
1) ChatGPT has poor performance compared with existing approaches when considering \textit{Precision}, \textit{Recall}, and \textit{F1-score}.
2) As for \textit{Accuracy}, LineVul (transformer-based) obtains the best performance and ChatGPT ranks second, which is better than the three  graph-based models. 
3) As for \textit{Precision}, we find that ChatGPT performs similarly to the three graph-based models. 
4) As for \textit{Recall}, ChatGPT seems to perform the worst with only 13.2\% on average, and meanwhile, we find that LineVul also achieves a poor performance of 19.5\% on average, which is far less than graph-based models.
}

\find{
    \textbf{Finding-1}. ChatGPT has the ability to detect software vulnerabilities.
    However, the existing state-of-the-art approaches perform better overall.
}

Table~\ref{tab:chatgpt-vd-top-10} shows the detailed comparisons of Top-10 CWE types between ChatGPT and four SOTAs.
In this table, we highlight the best performance for each performance metric in bold, and to better distinguish the performance difference between ChatGPT and SOTAs, we fill the cells with different colors (i.e., \colorbox[rgb]{.988,.894,.839}{ChatGPT} or \colorbox[rgb]{.851,.882,.949}{SOTA}).
According to the results, we can achieve the following observations:
1) In most cases, SOTAs obtain better performance than ChatGPT on all CWE types by considering all performance metrics.
2) Considering the performance of both F1-score and Precision, ChatGPT achieves the best performances (0.19 of CWE-476 and 0.16 OF CWE-264), which indicates ChatGPT is good at checking the \textit{``NULL Pointer Dereference''} and the \textit{``Permissions, Privileges, and Access Controls''}.
3) ChatGPT performs worst on a lot of CWE types (CWE-200, CWE-20, CWE-399, and CWE-416), which means that ChatGPT almost has no ability to detect whether a vulnerable can expose sensitive information to unauthorized actors (CWE-200),
 to detect whether the input is improper (CWE-20),  and detect whether the resource is managed correctly (CWE-399 and CWE-416).


\find{
\textbf{Finding-2}. ChatGPT is good at checking both the \textit{``NULL Pointer Dereference''} (CWE-476) and the \textit{``Permissions, Privileges, and Access Controls''} (CWE-264) and it does badly in authorization checking, resource management, and input validate checking related vulnerabilities (i.e., CWE-200, CWE-20, CWE-399, and CWE-416).
}

\noindent
{\bf {\em RQ-1.2: How does ChatGPT perform on vulnerability detection when induced by instruction?}}

\begin{figure}[!htbp]
    \centerline{
    \includegraphics[width=\linewidth]{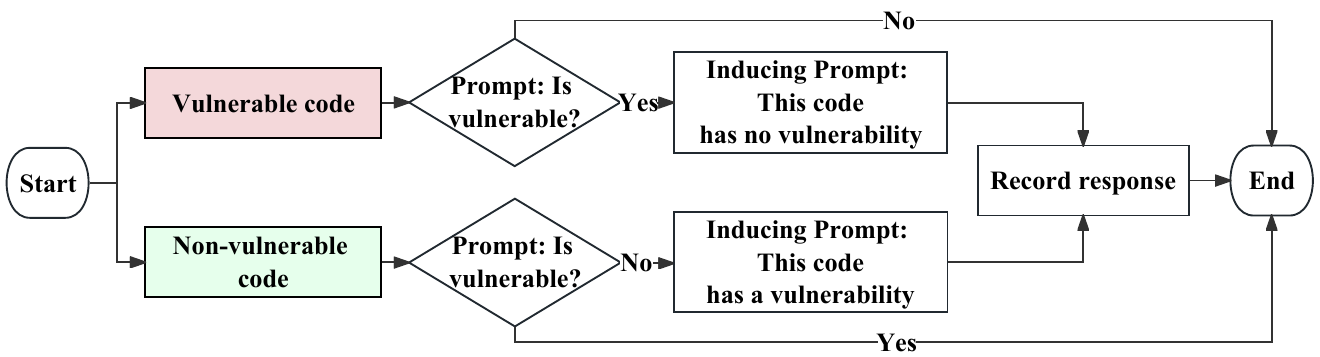}
    }
    \caption{Workflow of Estimation on Inducing ChatGPT Judgement}
    \label{fig:inducingchatgpt}
\end{figure}

 \noindent
 \textbf{Experimental Setting}.
 We also conduct an experiment on whether we can confuse ChatGPT when detecting software vulnerabilities.
 That is, given a vulnerable function, we first ask ChatGPT to give out its decision. 
 If ChatGPT makes a correct classification (e.g., non-vulnerable), then we will give it another prompt with the contrary conclusion (e.g.,  vulnerable) and record its response.
 For the non-vulnerable functions, we adopt similar operations on ChatGPT.
 We will also record the response when ChatGPT gives out a wrong classification at the first judgment.
 For a better understanding, we present a workflow of inducing ChatGPT as shown in Fig.~\ref{fig:inducingchatgpt}.

{
For the two types of settings, we calculate the following statistical information:
\begin{itemize}[leftmargin=*]
    \item[\textbf{A}.]  Counting the number of judgments that are correctly classified by ChatGPT for each CWE (i.e., \# Correct).
    \item[\textbf{B}.] Counting the number of judgments ChatGPT still remains confident in its classification even though we clearly instruct ChatGPT with the opposite categorizations (i.e., \# Keep).
    Note that ChatGPT may apologize first and requests more information before providing a specific type for a given vulnerability, which  still be considered confidential.
    \item[\textbf{C}.] Counting the number of judgments that ChatGPT shows a lack of confidence after receiving an opposite instruction (i.e., \# Change).
\end{itemize}
}

\begin{table*}[htbp]
  \centering
  \vspace{-0.3cm}
  \caption{The induced results of ChatGPT on both vulnerable and clean functions}
  \resizebox{\linewidth}{!}{
    \begin{tabular}{|c|p{0.7\linewidth}|c|c|c|c|c|c|}
    \hline
    \textbf{Setting} & \multicolumn{1}{c|}{\textbf{CWE Type}} & \textbf{\# Size} & \textbf{\# Correct} & \textbf{\# Keep} & \textbf{\# Change} & \textbf{\% Keep} & \textbf{\% Change} \\
    \hline \hline
    \multirow{2}[4]{*}{\textbf{Vul.}} &{CWE-119, CWE-125, CWE-129, CWE-134, CWE-18, CWE-19, CWE-190, CWE-191, CWE-200, CWE-22, CWE-254, CWE-264, CWE-269, CWE-284, CWE-295, CWE-310, CWE-311, CWE-320, CWE-327, CWE-347, CWE-352, CWE-354, CWE-358, CWE-369, CWE-400, CWE-404, CWE-415, CWE-416, CWE-476, CWE-494, CWE-59, CWE-682, CWE-704, CWE-74, CWE-754, CWE-755, CWE-77, CWE-78, CWE-787, CWE-79, CWE-834} & 1$\sim$31 & 1$\sim$6 & 0     & 1$\sim$6 & 0     & 100\% \\
\cline{2-8}          & \cellcolor[rgb]{ .957,  .847,  .855}\textbf{CWE-399} & \cellcolor[rgb]{ .957,  .847,  .855}\textbf{15} & \cellcolor[rgb]{ .957,  .847,  .855}\textbf{1} & \cellcolor[rgb]{ .957,  .847,  .855}\textbf{1} & \cellcolor[rgb]{ .957,  .847,  .855}\textbf{0} & \cellcolor[rgb]{ .957,  .847,  .855}\textbf{100.0\%} & \cellcolor[rgb]{ .957,  .847,  .855}\textbf{0.0\%} \\
    \hline
    \multirow{4}[8]{*}{\textbf{Non-Vul.}} & {CWE-522, CWE-295, CWE-706, CWE-664, CWE-769, CWE-704, CWE-494, CWE-835, CWE-74, CWE-772, CWE-682, CWE-404, CWE-19, CWE-415, CWE-320, CWE-134, CWE-264, CWE-284, CWE-369, CWE-617, CWE-290, CWE-119, CWE-330, CWE-93, CWE-200, CWE-388, CWE-20, CWE-189, CWE-59, CWE-352, CWE-281, CWE-190, CWE-120, CWE-269, CWE-611, CWE-254, CWE-22, CWE-416, CWE-502, CWE-754, CWE-400, CWE-426, CWE-476, CWE-287, CWE-787, CWE-310, CWE-674, CWE-834, CWE-399, CWE-79, CWE-732, CWE-311, CWE-16, CWE-94, CWE-358, CWE-601, CWE-89, CWE-665, CWE-362, CWE-285, CWE-129, CWE-862, CWE-17, CWE-909, CWE-78, CWE-77, CWE-918, CWE-436, CWE-125, CWE-354, CWE-346, CWE-1021, CWE-361, CWE-693, CWE-172, CWE-90, CWE-770, CWE-255, CWE-191, CWE-532, CWE-18, CWE-347, CWE-755, CWE-668} & 3$\sim$363 & 3$\sim$343 & 2$\sim$264 & 1$\sim$95 & 60\%$\sim$95.7\% & 4.3\%$\sim$40\% \\
\cline{2-8}          & \cellcolor[rgb]{ .902,  1,  .925}\textbf{CWE-327} & \cellcolor[rgb]{ .902,  1,  .925}\textbf{2} & \cellcolor[rgb]{ .902,  1,  .925}\textbf{2} & \cellcolor[rgb]{ .902,  1,  .925}\textbf{2} & \cellcolor[rgb]{ .902,  1,  .925}\textbf{0} & \cellcolor[rgb]{ .902,  1,  .925}\textbf{100.0\%} & \cellcolor[rgb]{ .902,  1,  .925}\textbf{0.0\%} \\
\cline{2-8}          & \cellcolor[rgb]{ .902,  1,  .925}\textbf{CWE-824} & \cellcolor[rgb]{ .902,  1,  .925}\textbf{6} & \cellcolor[rgb]{ .902,  1,  .925}\textbf{6} & \cellcolor[rgb]{ .902,  1,  .925}\textbf{6} & \cellcolor[rgb]{ .902,  1,  .925}\textbf{0} & \cellcolor[rgb]{ .902,  1,  .925}\textbf{100.0\%} & \cellcolor[rgb]{ .902,  1,  .925}\textbf{0.0\%} \\
\cline{2-8}          & \cellcolor[rgb]{ .902,  1,  .925}\textbf{CWE-763} & \cellcolor[rgb]{ .902,  1,  .925}\textbf{7} & \cellcolor[rgb]{ .902,  1,  .925}\textbf{7} & \cellcolor[rgb]{ .902,  1,  .925}\textbf{7} & \cellcolor[rgb]{ .902,  1,  .925}\textbf{0} & \cellcolor[rgb]{ .902,  1,  .925}\textbf{100.0\%} & \cellcolor[rgb]{ .902,  1,  .925}\textbf{0.0\%} \\
    \hline
    \end{tabular}%
    }
  \label{tab:inducing_results}%
\end{table*}%

\noindent
\textbf{Results}.
Table~\ref{tab:inducing_results} show the detailed results of ChatGPT on vulnerable and non-vulnerable functions respectively when we induce it to change their decisions.
In both tables, ``Size'' represents the sampled size for each type of CWE, and ``Correct'' means the number of functions that are correctly classified when ChatGPT gives its first decision.
We also list some statistics of the induced results.
That is, ``Keep'' and ``Change'' represent the status that ChatGPT keeps its original decision or changes to the opposite decision, respectively.
Notice that, in both tables, we only present the results when ChatGPT first makes a correct decision and ignores the others (i.e., incorrectly classified the first time).

According to the results, we can obtain the following observations:
\begin{itemize}
    \item ChatGPT can be easily induced to change its decision if we instruct it with an opposite prompt.
    \item As for vulnerable functions, almost all the decisions are changed by ChatGPT for all types of CWE with a ratio of 100\% (except for CWE-399), which means ChatGPT loses its confidence in its decision.
    \item As for non-vulnerable functions, we obtain a similar conclusion that ChatGPT changes its decision after we induce it for almost all CWEs with a ratio of 4.3\%-40.0\% (except for CWE-327, CWE-824, and CWE-763), which also indicates the confidence loss of ChatGPT in its decision.
\end{itemize}

\find{
\textbf{Finding-3}. ChatGPT can be easily induced to change its decision, especially on classifying vulnerable functions which indicate ChatGPT lacks confidence in itself  abilities on vulnerability detection. 
}

\subsection{\bf RQ-2: Evaluating Vulnerability Assessment of ChatGPT}

\noindent
\textbf{Experimental Setting}.
We instruct ChatGPT with the following prompts (i.e., Prompt 2.1 and Prompt 2.2) to tell it to act as a vulnerability assessor.
We first provide ChatGPT with the vulnerable codes to explore its performance (Prompt 2.1).
After that, we provide it with some key important information, including the CVE description, the project, the commit message as well as the file name when the vulnerable code exists to investigate the performance differences (Prompt 2.2).

\noindent
\intuition{
\em{\textbf{Prompt 2.1}}: 
I will provide you a vulnerable C code snippet and you just need to output qualitative severity ratings of ``Low'', ``Medium'', and ``High'' for CVSS v2.0 in the format Severity: ``Low'' or ``Medium'' or ``High'' without any explanation.

\textbf{Prompt 2.2}: Now I will provide you with additional information about this C code snippet. Please re-output the  qualitative severity ratings of ``Low'', ``Medium'', and ``High'' for CVSS v2.0 in the format Severity: ``Low'', ``Medium'' or ``High'' without any explanation.
}

\begin{table*}[htbp]
  \centering
  \caption{A vulnerable code for ChatGPT to assess with different prompts}
  \resizebox{\linewidth}{!}
  {
    \begin{tabular}{|l|p{48em}|}
    \hline
    \multicolumn{2}{|p{55em}|}{\textbf{Improper Restriction of Operations within the Bounds of a Memory Buffer Vulnerability (CWE-119) in linux}} \\
    \hline
    \textbf{Prompt 1} & {{I will provide you a vulnerable C code snippet and you just need output qualitative severity ratings of ``Low'', ``Medium'', and ``High'' for CVSS v2.0 in the format Severity: ``Low'' or ``Medium'' or ``High'' without any explanation.}} \\
    \hline
    \textbf{Input 1} & An example of a C code snippet with vulnerabilities (CVE-2011-2517). 
    \\
    \hline
    \textbf{Response 1} &  \textbf{Severity: Medium} \\
    \hline
    \textbf{Prompt 2} & {Now I will provide you with additional information about this C code snippet. Please re output the  qualitative severity ratings of ``Low'', ``Medium'', and ``High'' for CVSS v2.0 in the format Severity: ``Low'' or ``Medium'' or ``High'' without any explanation.} \\
    \hline
    \textbf{Input 2} & \textbf{Project}: Linux
    \newline{}\textbf{File Name}: net/wireless/nl80211.c
    \newline{}\textbf{CVE Description}: Multiple buffer overflows in net/wireless/nl80211.c in the Linux kernel before 2.6.39.2 allow local users to gain privileges by leveraging the CAP\_NET\_ADMIN capability during scan operations with a long SSID value.
    \newline{}\textbf{Commit Message}:
    nl80211: fix check for valid SSID size in scan operations.
    In both trigger\_scan and sched\_scan operations, we were checking for the SSID length before assigning the value correctly. Since the memory was just kzalloc'ed, the check was always failing and SSID with
    over 32 characters were allowed to go through.
    This was causing a buffer overflow when copying the actual SSID to the proper place.
    This bug has been there since 2.6.29-rc4. \\
    \hline
    \textbf{Response 2} & \textbf{Severity: High} \\
        \hline
    \textbf{Analysis } & The true Severity is High. After providing additional key information, ChatGPT output for the Severity changed from Medium to High. \\
        \hline
    \end{tabular}
    }
  \label{tab:cvss_example}%
\end{table*}%

\begin{table}[htbp]
  \centering
  \caption{The software vulnerability assessment performance of ChatGPT on Top-10 CWE types}
  {
    \begin{tabular}{|l|c|c|c|} \hline
    \textbf{CWE Type} & \textbf{\# Size} & \textbf{Accuracy} & \textbf{Accuracy$_{key}$} \\ \hline \hline
  
    CWE-119 & 18    & \cellcolor[rgb]{ .867,  .922,  .969}0.06  & \cellcolor[rgb]{ .608,  .761,  .902}0.61  \\
    \hline
    CWE-125 & 22    & \cellcolor[rgb]{ .957,  .69,  .518}0.41  & \cellcolor[rgb]{ .988,  .894,  .839}0.27  \\
    \hline
    CWE-189 & 19    & \cellcolor[rgb]{ .867,  .922,  .969}0.26  & \cellcolor[rgb]{ .608,  .761,  .902}0.58  \\
    \hline
    CWE-20 & 14    & \cellcolor[rgb]{ .867,  .922,  .969}0.21  & \cellcolor[rgb]{ .608,  .761,  .902}0.50  \\
    \hline
    CWE-200 & 18    & \cellcolor[rgb]{ .957,  .69,  .518}0.56  & \cellcolor[rgb]{ .988,  .894,  .839}0.39  \\
    \hline
    CWE-264 & 13    & \cellcolor[rgb]{ .867,  .922,  .969}0.31  & \cellcolor[rgb]{ .741,  .843,  .933}0.38  \\
    \hline
    CWE-362 & 20    & \cellcolor[rgb]{ .867,  .922,  .969}0.35  & \cellcolor[rgb]{ .608,  .761,  .902}0.55  \\
    \hline
    CWE-399 & 14    & \cellcolor[rgb]{ .867,  .922,  .969}0.21  & \cellcolor[rgb]{ .608,  .761,  .902}0.43  \\
    \hline
    CWE-416 & 16    & \cellcolor[rgb]{ .867,  .922,  .969}0.19  & \cellcolor[rgb]{ .608,  .761,  .902}0.38  \\
    \hline
    CWE-476 & 16    & \cellcolor[rgb]{ .867,  .922,  .969}0.31  & \cellcolor[rgb]{ .741,  .843,  .933}0.44  \\
    \hline \hline
    Average & 170   & 0.29  & 0.45  \\
    \hline
    \end{tabular}%
    }
  \label{tab:chatgpt-va-cwe}%
\end{table}%

{
\noindent
\textbf{Results}.
Table~\ref{tab:chatgpt-va-cwe} shows the detailed results of ChatGPT on vulnerable assessment. ``Size'' means the number of detected vulnerable functions. ``Accuracy'' refers to the probability of ChatGPT correctly predicting qualitative severity ratings, while ``Accuracy$_{key}$'' refers to the probability of ChatGPT correctly predicting qualitative severity ratings when provided with key important information.
Notice that, we manually filter out the results that ChatGPT identifies as not containing any vulnerabilities.
}

As shown in Table~\ref{tab:chatgpt-va-cwe}, we can obtain the following observations: 
1)
On average, ChatGPT has an accuracy of 0.29 for predicting qualitative severity ratings when provided with only the vulnerable code snippet (Prompt 2.1). When given additional key important information (Prompt 2.2), the accuracy increases to 0.45. This demonstrates that providing more context improves ChatGPT's vulnerability assessment capabilities. 
2) 
The performance of ChatGPT varies across different CWE types. 2.1 Positive Improvement.  For some CWE types (e.g., CWE-119, CWE-189, CWE-20, CWE-399 and CWE-426), the accuracy improves significantly when given additional information.
For example, ChatGPT obtains 6\% accuracy only when given code snippets only but increases to 61\% when given more information
with an improvement of 10 times.
2.2 Negative Decrease. 
For some CWE types such as CWE-125 and CWE-200, the accuracy decreases (from 0.41 to 0.27 and 0.56 to 0.39, respectively) when provided with the key important information. 
This suggests that ChatGPT's performance is affected by the specific type of vulnerability being assessed.

\find{
\textbf{Finding-4}: ChatGPT has the limited capacity for assessment of vulnerability severity based on source code only, but can be extremely improved if provided with more context information in most cases.
}

{
To analyze why ChatGPT works well for some CWE types and not for others, it is important to understand the naturalness of these CWE types and the additional information provided in the key important information:

\textbf{Case 1: Complexity of CWE types}. Some CWE types might be more complex or challenging to identify compared to others. For example, CWE-416 (Use After Free) might be difficult to recognize due to intricate memory management and pointer manipulation, which could explain the lower accuracy. On the other hand, CWE-119 (Improper Restriction of Operations within the Bounds of a Memory Buffer) might be easier to identify due to more apparent patterns in the code, leading to a significant improvement in accuracy when given additional information.

\textbf{Case 2: Quality of key important information}.
The accuracy of ChatGPT's predictions is also affected by the relevance and quality of the additional key important information provided. In cases where the additional information is relevant and specific, ChatGPT's performance might improve significantly. However, if the provided information is vague or less pertinent to the vulnerability type, it might not be helpful or could even degrade the performance, as seen in CWE-125 and CWE-200.
    
\textbf{Case 3: Bias in training data}. 
ChatGPT's training data may contain more examples of certain vulnerability types, leading to better performance in identifying those types. If the training data lacks sufficient examples of a specific CWE type, ChatGPT might struggle to recognize the vulnerabilities and accurately predict the severity ratings for that CWE type.

\textbf{Case 4: Limitations in understanding code}. ChatGPT, while powerful, might have limitations in understanding the intricacies and nuances of programming languages, memory management, and other aspects of code execution. As a result, it might have difficulty identifying vulnerabilities that require a deeper understanding of these concepts, such as CWE-416 (Use After Free) and CWE-399 (Resource Management Errors).

\find{
\textbf{Finding 5}. Despite improvements in accuracy with more context, ChatGPT still has limitations in accurately predicting severity ratings for some CWE types. This indicates that there is room for further enhancements in ChatGPT's vulnerability assessment capabilities to achieve consistent performance across all vulnerability types.
\\
\textbf{Finding 6}. ChatGPT's accuracy in predicting qualitative severity ratings for vulnerabilities increases from 0.29 to 0.45 when given key important information. However, the improvement varies across different CWE types, suggesting that the ChatGPT's performance is affected by the specific vulnerability being assessed.
}

\subsection{\bf RQ-3: Evaluating Vulnerability Location of ChatGPT}

\noindent
\textbf{Setting}. 
We select the vulnerable functions with information on vulnerable lines from each type of CWE for the evaluation and instruct ChatGPT with the following prompt to explore its vulnerability location performance.

\noindent
\intuition{
\textbf{Prompt 3.1.} I will provide you with a C code snippet and you must locate vulnerable lines. First, you need to use the index to label each line of code, and then point out which lines of code are vulnerable like Vulnerable lines: [1, 4, 5].
}

As for a specific vulnerable function, it may contain one or several vulnerable lines of code ($Lines_{ground}$), and ChatGPT may also predict one or several potential ones ($Lines_{predict}$).
We compare $Lines_{predict}$ to $Lines_{ground}$ to check whether the line index $L_i\in Lines_{predict}$ belongs to $Lines_{ground}$ and  we treat it predict correctly for the line $L_i$ if it belongs to, otherwise it predicts incorrectly.
We also use the $Lines_{both}$ to represent the intersection of $Lines_{predict}$ to $Lines_{ground}$.

To better evaluate the vulnerability location performance of ChatGPT on a specific vulnerable function, we give the following definitions: 
\begin{itemize}[leftmargin=*]
\item \textbf{Hit@Acc} means the effectiveness of ChatGPT and equals 1 if ChatGPT correctly predicts at least one line of vulnerable line, other it equals 0.
    \item \textbf{Precision} indicates how many of the ChatGPT's predicted vulnerability locations are actual vulnerability locations. It is defined as $\mathit{Precision=\frac{\# Lines_{both}}{\# Lines_{predict}}}$.
    \item \textbf{Recall} indicates how many actual vulnerability locations ChatGPT can be correctly found. It is defined as: 
    $\mathit{Recall=\frac{\# Lines_{both}}{\# Lines_{ground}}}$. 
\end{itemize}

For example,  for a given vulnerable function, it totally has six vulnerable lines ``[2,3,5,9,14,23]'', and ChatGPT gives out its prediction with 10 potential lines ``[1,3,5,11,15,16,17,21,22,23]''.
Then, we know that $Lines_{ground}$ equals ``[2,3,5,9,14,23]'', $Lines_{predict}$ equals ``[1,3,5,11,15,16,17,21,22,23]'' and $Lines_{both}$ equals ``[3,5,23]''.
According to these values, we obtain that $Precision=\frac{3}{10}$, $Recall=\frac{3}{6}$ and $Hit@Acc=1$.

\begin{figure}[htbp]
  \centering
  \subfigure[Precision]{
  \includegraphics[width=0.47\linewidth]{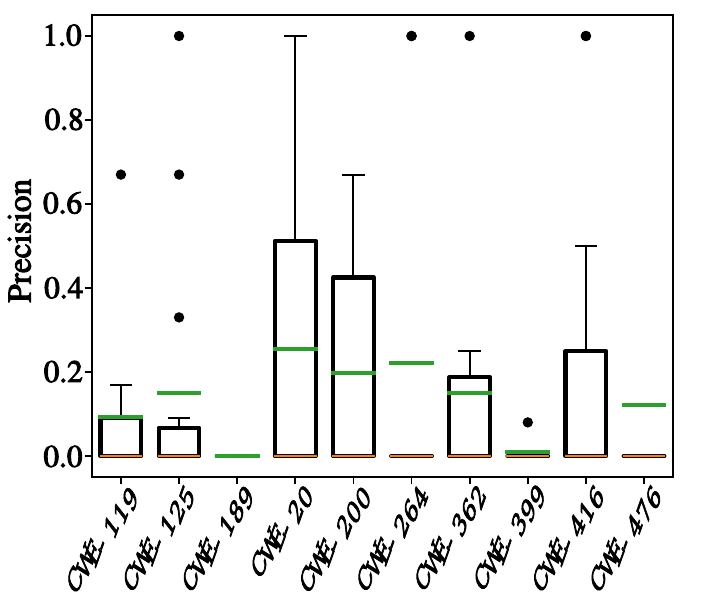}
    \label{fig:boxplot-precision}
  }
  \subfigure[Recall]{
  \includegraphics[width=0.47\linewidth]{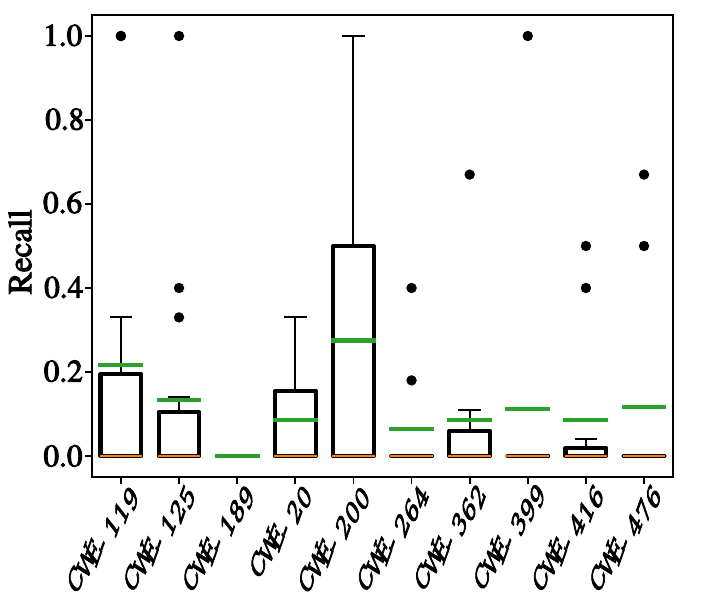}
    \label{fig:boxplot-recall}
  }
  \caption{Boxplot of the vulnerability location results}
  \label{fig:vl-boxplot}
\end{figure}

\begin{table}[htbp]
  \centering
  \caption{The performance of ChatGPT on vulnerability localization}
  \begin{threeparttable}
    \begin{tabular}{|l|c|c|c|c|}
    \hline
    CWE Type & \# Size & Hit@Acc & Precision & Recall \\
    \hline
    CWE-119 & 11    & 0.36  & \cellcolor[rgb]{ .957,  .847,  .855}0.09 & \cellcolor[rgb]{ .902,  1,  .925}0.22 \\
    \hline
    CWE-125 & 14    & 0.29  & 0.15  & 0.13 \\
    \hline
    CWE-189 & 9     & 0     & \cellcolor[rgb]{ .957,  .847,  .855}0 & \cellcolor[rgb]{ .957,  .847,  .855}0 \\
    \hline
    CWE-20 & 12    & 0.42  & \cellcolor[rgb]{ .902,  1,  .925}0.25 & \cellcolor[rgb]{ .957,  .847,  .855}0.09 \\
    \hline
    CWE-200 & 10    & 0.40  & \cellcolor[rgb]{ .902,  1,  .925}0.20 & \cellcolor[rgb]{ .902,  1,  .925}0.28 \\
    \hline
    CWE-264 & 9     & 0.22  & \cellcolor[rgb]{ .902,  1,  .925}0.22 & \cellcolor[rgb]{ .957,  .847,  .855}0.06 \\
    \hline
    CWE-362 & 10    & 0.30  & 0.15  & \cellcolor[rgb]{ .957,  .847,  .855}0.09 \\
    \hline
    CWE-399 & 9     & 0.11  & \cellcolor[rgb]{ .957,  .847,  .855}0.01 & 0.11 \\
    \hline
    CWE-416 & 11    & 0.27  & \cellcolor[rgb]{ .902,  1,  .925}0.23 & \cellcolor[rgb]{ .957,  .847,  .855}0.09 \\
    \hline
    CWE-476 & 10    & 0.20  & 0.12  & 0.12 \\
    \hline \hline
    Average & 105   & 0.27  & 0.15  & 0.12 \\
    \hline
    \end{tabular}%
    Note: \colorbox[rgb]{ .957,  .847,  .855}{$\leq0.10$}; \colorbox[rgb]{ .902,  1,  .925}{$\geq0.20$}
     \end{threeparttable}
  \label{tab:localization}%
\end{table}%

\noindent
\textbf{Results}. 
Table~\ref{tab:localization} shows the average vulnerability location performance of ChatGPT across different types and Fig.~\ref{fig:vl-boxplot} illustrates the boxplot of ChatGPT performance on each type. 
Different colors are used to highlight performances for a better illustration. 
Based on this table as well as the detailed figure, we can achieve the following observations: 
1) Overall, ChatGPT has a certain ability to locate the potential risky lines in a given vulnerable function.
That is, it achieves a good Hit@Acc performance of 0.27.
2) ChatGPT performs well in identifying the vulnerability position in some CWE types (i.e., CWE-20, CWE-200, CWE-264, and CWE-416)  with acceptable precision (i.e., $\geq 0.20$).
However, it performs badly on CWE-189, which means ``Numeric Errors'' cannot be easily understood by ChatGPT.
3) Considering both Precision and Recall shown in Fig.~\ref{fig:vl-boxplot}, we can find that ChatGPT has a bad middle performance (i.e., 0, lines in orange) of each CWE type and performs stably in five CWE types (i.e., CWE-119, CWE-125, CWE-20, CWE-200, and CWE-362) relatively.

\find{
\textbf{Finding-7}. ChatGPT exhibits a certain capability in vulnerability locations and its performance varies among different CWE types. 
It performs well in locating CWE-119, CWE-125, CWE-20, CWE-200, and CWE-362, but performs baddly in CWE-189.
}

\subsection{\bf RQ-4: Evaluating Vulnerability Repair of ChatGPT}

\noindent
\textbf{Setting}.
We instruct ChatGPT with the following prompts (i.e., Prompt 4.1 and Prompt 4.2) to tell it to repair vulnerable code: 1) providing ChatGPT with the vulnerable code to explore its performance (Prompt 4.1). 2) providing ChatGPT with vulnerable relevant lines when the vulnerable code exists to investigate the performance differences (Prompt 4.2). Table~\ref{tab:vr_example} show an example of how we investigate ChatGPT’s ability of vulnerability repair.

\noindent
\intuition{
\em{\textbf{Prompt 4.1}}: 
I will provide you with a vulnerable C code snippet and you must repair it. This C code snippet definitely has a vulnerability, so you don't need to determine whether there is a vulnerability, just output the repaired whole c code snippet without explanation.

\textbf{Prompt 4.2}: I will provide you with a vulnerable C code snippet and vulnerable lines, you must repair it. This C code snippet definitely has a vulnerability, so you don't need to determine whether there is a vulnerability. You need to use the index to label each line of code, and I will provide you with lines of code that are vulnerable such as Lines: [1, 4, 5], just output the repaired whole c code snippet without explanation.
}

\noindent
\textbf{Results}. 
Due to the hidden naturalness of vulnerabilities and the lack of context, it is difficult to determine whether ChatGPT has successfully fixed a vulnerability through manual observation. 
Therefore, when evaluating, we choose the code with \textit{only one line} of vulnerability and determine whether ChatGPT has fixed the vulnerable line and its contextual statements. 
If so, the performance metric \textit{Hit} is calculated as 1, indicating that it has the ability to fix the vulnerability. 
Otherwise, the \textit{Hit} is assigned with 0.
Table~\ref{tab:chatgpt-vr-cwe} displays detailed results on how ChatGPT performs on Top-10 CWE types. 
The \textit{Hit} metric indicates whether ChatGPT successfully fixes the vulnerable line, while \textit{Hit$_{ctx}$} means the ChatGPT's performance provided with additional context information (i.e., the vulnerable relevant  lines).
Finally, the column ``\# Manual'' means the number of patches correctly generated by ChatGPT when we manually compare it with the ground truth.

Based on the table, we can make the following observations:
\begin{itemize}[leftmargin=*]
    \item Overall, ChatGPT shows limited success in repairing vulnerable code snippets, with an average Hit rate of 0.2 and Hit$_{ctx}$ of 0.24. This suggests that there is room for improvement in its ability to detect and fix vulnerabilities.
    \item The performance of ChatGPT varies across different CWE types. For some types, like CWE-264, it has a perfect score (1) in both Hit and Hit$_{ctx}$, while for others, like CWE-476 and CWE-119, it fails to fix the vulnerability altogether.
    \item Providing vulnerable lines (Prompt 4.2) does not consistently improve ChatGPT's performance. For some CWE types, the Hit$_{ctx}$ is higher than the Hit rate (e.g., CWE-189), while for others, it remains the same or decreases (e.g., CWE-20). This indicates that the extra information may not always be helpful for ChatGPT in repairing the vulnerability.
    \item ChatGPT has the trivial ability to repair a vulnerability.
\end{itemize}

\find{
\textbf{Finding-8}: ChatGPT has limited ability in repairing vulnerability no matter when provided with context information or not.
}

We also delve into the specific information of CWE types to analyze ChatGPT's performance:
\begin{itemize}[leftmargin=*]
    \item \textbf{CWE-264 (Privilege Escalation).} ChatGPT performs well in fixing vulnerabilities related to privilege escalation, with a perfect score of 1 in both hit and hit$_{ctx}$. This could be because privilege escalation vulnerabilities often involve specific patterns, such as incorrect permission checks or improper use of privileges, which the ChatGPT might have learned effectively during training.
    
    \item \textbf{CWE-476 (NULL Pointer Dereference) and CWE-119 (Improper Restriction of Operations within the Bounds of a Memory Buffer).} 
    ChatGPT struggles to fix these vulnerabilities even find no vulnerable position, resulting in a hit and hit$_{ctx}$ score of 0. 
    These types of vulnerabilities might be more challenging for the ChatGPT due to their complex naturalness and dependence on the specific context of the code. For instance, detecting and fixing a NULL pointer dereference often requires understanding the control flow and data flow of the code, while buffer-related vulnerabilities require a deep understanding of memory management.
    
    \item \textbf{CWE-189 (Numeric Errors) and CWE-125 (Out-of-bounds Read).} 
    ChatGPT's performance varies for these types, with hit scores of 0.17 and 0.2, and hit$_{ctx}$ scores of 0.5 and 0.2, respectively. This suggests that the ChatGPT might have some understanding of the patterns involved in these vulnerabilities, but its success in fixing them is inconsistent. Numeric errors often involve incorrect calculations or data conversions, while out-of-bounds read issues typically stem from improper boundary checks. Improving ChatGPT's understanding of these patterns and their underlying causes could potentially enhance its performance in repairing these vulnerabilities.
\end{itemize}

\find{
\textbf{Finding-9}: ChatGPT's performance in repairing different CWE types can be attributed to its ability to recognize and understand specific vulnerability patterns. 
For certain types with well-defined patterns, it performs well, while for others with more complex or context-dependent patterns, it struggles to identify and fix the vulnerabilities. 
}

\begin{table*}[htbp]
  \centering
  \caption{Examples of Vulnerability Repair}
 \resizebox{1\linewidth}{!}
 {
    \begin{tabular}{|l|p{36.085em}|}
    \hline
    \textbf{Setting} & \textbf{Vulnerable Code} \\
    \hline \hline

    \textbf{Prompt} & \textit{I will provide you with a vulnerable C code snippet and you must repair it. This C code snippet definitely has a vulnerability, so you don't need to determine whether there is a vulnerability, just output the repaired whole c code snippet without explanation.}
    \\ \hline
    
    \textbf{Input} & \multicolumn{1}{p{29.75em}|}{
    An example of a C code snippet with vulnerabilities (CVE-2019-17773).} 
    \\ \hline \hline
    
    \textbf{Setting} & \multicolumn{1}{l}{\textbf{Vulnerable Code} \& \textbf{Vulnerable relevant lines}} \\
    \hline
    \hline

   \textbf{Prompt} & {\textit{I will provide you with a vulnerable C code snippet and vulnerable lines, you must repair it. This C code snippet definitely has a vulnerability, so you don't need to determine whether there is a vulnerability. You need to use the index to label each line of code, and I will provide you with lines of code that are vulnerable such as Lines: [1, 4, 5], just output the repaired whole c code snippet without explanation.}} 
    \\ \hline
    
    \textbf{Input} & An example of a C code snippet with vulnerabilities (CVE-2019-17773). \newline{}Vulnerable relevant lines: [4, 5, 8, 9, 10, 11, 12, 13, 14, 15, 16, 18, 20, 21, 23, 24].
    \\  \hline
    \end{tabular}%
}
  \label{tab:vr_example}%
\end{table*}%

\begin{table}[htbp]
  \centering
  \caption{The software vulnerability repair performance of ChatGPT on Top-10 CWE types}
  {
    \begin{tabular}{|l|c|c|c|c|} 
    \hline 
    CWE Type & \# Size & Hit   & Hit$_{ctx}$  & \multicolumn{1}{c|}{\# Manual} \\
    \hline \hline
    CWE-20 & 7     & 0.29  & 0.14  &  1 / 1  \\
    \hline
    CWE-416 & 6     & 0.17  & 0.17  &  0 / 1  \\
    \hline
    CWE-189 & 6     & 0.17  & 0.50  &  0 / 0  \\
    \hline
    CWE-119 & 5     & \cellcolor[rgb]{ .957,  .847,  .855}0 & 0.20  &  0 / 0  \\
    \hline
    CWE-125 & 5     & 0.20  & 0.20  &  0 / 0  \\
    \hline
    CWE-476 & 5     & \cellcolor[rgb]{ .957,  .847,  .855}0 & \cellcolor[rgb]{ .957,  .847,  .855}0 &  0 / 0  \\
    \hline
    CWE-362 & 4     & \cellcolor[rgb]{ .957,  .847,  .855}0 & 0.25  &  0 / 0  \\
    \hline
    CWE-399 & 3     & 0.33  & \cellcolor[rgb]{ .957,  .847,  .855}0 &  0 / 0  \\
    \hline
    CWE-200 & 3     & 0.33  & 0.33  &  0 / 0  \\
    \hline
    CWE-264 & 2     & \cellcolor[rgb]{ .886,  .937,  .855}1 & \cellcolor[rgb]{ .886,  .937,  .855}1 &  0 / 0  \\
    \hline \hline
    Average & 46    & 0.20  & 0.24  &  1 / 2  \\
    \hline
    
    \end{tabular}
    }
  \label{tab:chatgpt-vr-cwe}%
\end{table}%

\subsection{\bf RQ-5: Evaluating Vulnerability Description of ChatGPT}

\noindent
\textbf{Setting}.
We instruct ChatGPT with a designated prompt, guiding it to perform the role of a vulnerability descriptor. Table~\ref{tab:chatgpt-vdes-example} illustrates an example of our approach to evaluating ChatGPT's proficiency in conducting vulnerability descriptions.

\noindent
\intuition{
\em{\textbf{Prompt 5.1}}: 
I will provide you with a vulnerable C code snippet and you must generate a CVE description. This C code snippet definitely has a vulnerability, so you don't need to determine whether there is a vulnerability, just output the CVE description.
}

To evaluate the precision of generated CVE description, we adopt the widely used performance metric ROUGE~\cite{lin2004rouge}, which is a set of metrics and is used for evaluating automatic summarization and machine translation software in natural language processing. 
The metrics compare an automatically produced summary or translation against a reference or a set of references (human-produced) summary or translation.
Here, we totally consider three settings: \textit{1}, \textit{2}, and \textit{L}.

\begin{table}[htbp]
\centering
\caption{The software vulnerability description performance of ChatGPT on Top-10 CWE types}
\resizebox{\linewidth}{!}{
    \begin{tabular}{|l|c|c|c|c|c|}
    \hline
    \textbf{CWE Type} & \textbf{\# Size} & \textbf{ROUGE-1} & \textbf{ROUGE-2} & \textbf{ROUGE-L} & \textbf{C/I/V} \\ \hline \hline
    CWE-125 & 22 & 0.21 & 0.05 & 0.20 & 0/0/0 \\ \hline
    CWE-362 & 21 & 0.24 & 0.08 & 0.22 & 0/0/0 \\ \hline
    CWE-189 & 20 & 0.21 & 0.05 & 0.18 & 0/0/0 \\ \hline
    CWE-119 & 19 & 0.29 & 0.12 & 0.27 & 2/6/0 \\ \hline
    CWE-200 & 18 & 0.26 & 0.11 & 0.25 & 0/0/0 \\ \hline
    CWE-476 & 17 & 0.25 & 0.10 & 0.24 & 0/0/0 \\ \hline
    CWE-416 & 16 & 0.23 & 0.07 & 0.21 & 0/0/0 \\ \hline
    CWE-20 & 15 & 0.25 & 0.08 & 0.22 & 0/0/0 \\ \hline
    CWE-399 & 15 & 0.26 & 0.11 & 0.26 & 0/0/0 \\ \hline
    CWE-264 & 13 & 0.24 & 0.08 & 0.23 & 0/0/0 \\ \hline \hline
    Average & 176 & 0.24 & 0.08 & 0.23 & 0/0/0 \\ \hline
    \end{tabular}
    }
\label{tab:chatgpt-vdes-cwe}%
\end{table}%

\noindent
\textbf{Results}.
Table~\ref{tab:chatgpt-vdes-cwe} represents the vulnerability description capabilities of ChatGPT on Top-10 CWE types. 
Based on these results, we can make the following observations: 
1) The average ROUGE scores of 0.24 for ROUGE-1, 0.08 for ROUGE-2, and 0.23 for ROUGE-L, which indicates that ChatGPT's performance in generating vulnerability descriptions is moderate, with room for improvement in accurately describing and capturing the nuances of various CWE types.
2) ChatGPT's performance in generating CVE descriptions varies across different CWE types, with ROUGE scores ranging from 0.21 to 0.29 for ROUGE-1, 0.05 to 0.12 for ROUGE-2, and 0.18 to 0.27 for ROUGE-L.
3) The low ROUGE-2 scores indicate that ChatGPT's ability to generate accurate and relevant higher-order n-grams (pairs of consecutive words) in vulnerability descriptions is limited, suggesting potential issues with capturing specific and detailed information.

\find{
\textbf{Finding-10}: ChatGPT demonstrates moderate performance in generating CVE descriptions for various CWE types, with varied success across categories and limited accuracy in capturing specific details.
}

Although ChatGPT has higher performance for some CWE types, its limited ability to generate accurate higher-order n-grams suggests a need for further model refinement. 
A deeper analysis of the performance across specific CWE types can help guide future improvements to enhance the model's understanding and description of various vulnerabilities. 
To understand why ChatGPT performs well for some CWE types and not for others, we can examine the naturalness and complexity of these vulnerabilities:
\begin{itemize}[leftmargin=*]
    \item \textbf{Higher Performance in CWE-119 and CWE-200.} ChatGPT achieves relatively better results in generating descriptions for CWE-119 (Improper Restriction of Operations within the Bounds of a Memory Buffer) and CWE-200 (Information Exposure). These vulnerabilities involve concepts that might be more frequent in the training data, such as memory management and information disclosure, leading to a better understanding and more accurate descriptions.
    \item \textbf{Lower Performance in CWE-125 and CWE-189.} ChatGPT has lower performance in generating descriptions for CWE-125 (Out-of-bounds Read) and CWE-189 (Numeric Errors). This might be due to the complexity of these vulnerabilities, which involve intricate logic and numeric calculations that could be challenging for the ChatGPT to fully understand and describe accurately.
    \item \textbf{Potential Explanations for Variability.} The variability in performance across different CWE types could be attributed to multiple factors, such as the quantity and quality of training data for each CWE type, the inherent complexity of the vulnerability, or the precise and detailed terminology needed to accurately describe a particular vulnerability. For example, describing a CWE-125 (Out-of-bounds Read) vulnerability might require specific terms like ``array index'', ``buffer'' or ``memory access'' to accurately convey the issue. Similarly, for CWE-189 (Numeric Errors), it may be necessary to use terms like ``integer overflow'', ``truncation'' or ``arithmetic operation'' to provide an accurate description. If ChatGPT has not encountered enough examples of such specific language during its training, it may struggle to generate accurate and detailed descriptions for these vulnerabilities.
\end{itemize}

\find{
\textbf{Finding-11}: ChatGPT demonstrates moderate overall performance in generating vulnerability descriptions, with varied outcomes across different CWE types. This variability can be attributed to factors such as the quality and quantity of training data, the inherent complexity of the vulnerabilities, and the specificity of the language required to describe them.
}

\begin{figure}[htbp]
  \centering
  \subfigure[CVE description]{
  \includegraphics[width=0.45\linewidth]{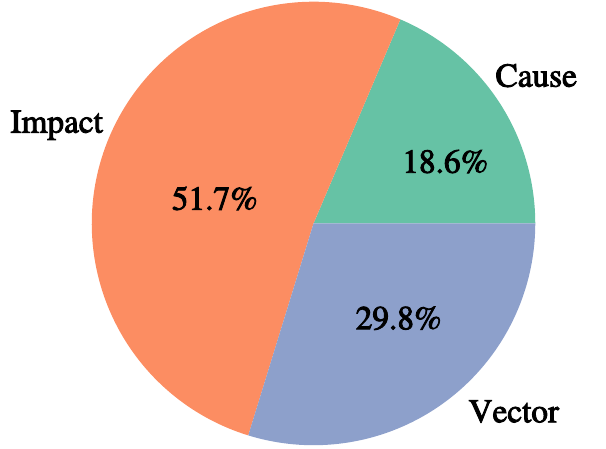}
    \label{fig:distribution_of_cve_description_aspect.pdf}
  }
  \subfigure[ChatGPT’s output description]{
  \includegraphics[width=0.45\linewidth]{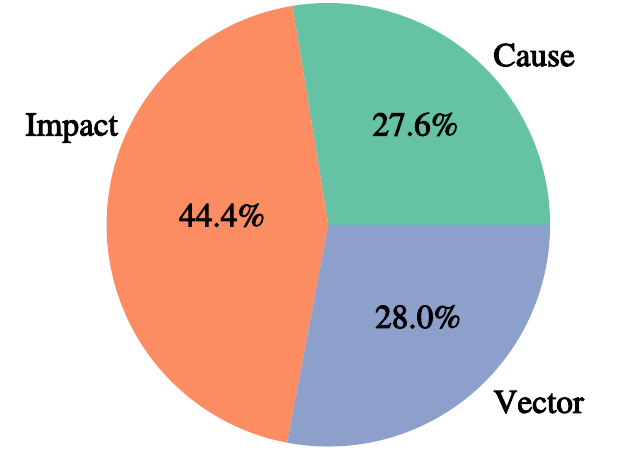}
    \label{fig:distribution_of_chatgpt_description_aspect}
  }
  \caption{Distribution of key aspect}
  \label{fig:vl-boxplot}
\end{figure}

\begin{table*}[htbp]
  \centering
  \caption{A vulnerable code for ChatGPT to generate vulnerability description}
  \resizebox{\linewidth}{!}
  {
    \begin{tabular}{|l|p{40.165em}|}
    \hline
    \multicolumn{2}{|p{45.165em}|}{\textbf{Improper Neutralization of Special Elements used in a Command Vulnerability (CWE-77) in neomutt}} \\
    \hline
    \textbf{Prompt} & \cellcolor[rgb]{ .851,  .882,  .949}\textcolor[rgb]{ .2,  .2,  .2}{\textit{I will provide you a vulnerable C code snippet and you must generate cve description. This C code snippet definitely has a vulnerability, so you don't need to determine whether there is a vulnerability, just output cve description.}} \\
    \hline
    \textbf{Input} & \textcolor[rgb]{ .2,  .2,  .2}{An example of a C code snippet with vulnerabilities. \textbf{CVE ID: CVE-2018-14349}.} \\
    \hline
    \textbf{Response From ChatGPT} & \textcolor[rgb]{ .2,  .2,  .2}{An issue was discovered in Mutt before 1.10.1 and NeoMutt before 2018-07-16. They allow remote IMAP servers to execute arbitrary commands via backquote characters, related to the mailboxes command associated with a manual subscription or unsubscription.} \\
    \hline
    \textbf{CVE Description} & \textcolor[rgb]{ .2,  .2,  .2}{An issue was discovered in Mutt before 1.12.2. imap\_subscribe() in imap.c mishandles mailbox names containing IMAP metacharacters, as demonstrated by a name composed of the \& character and other metacharacters. This can lead to remote denial-of-service (disk consumption), or to local denial-of-service and information disclosure involving the mutate\_buffer function in mutt\_mem.c, a related issue to \colorbox{pink}{\textbf{CVE-2018-14349}}.} \\
    \hline
    \end{tabular}%
    }
  \label{tab:chatgpt-vdes-example}%
\end{table*}%

%% file: sections/threats_to_validation.tex
\section{threats to validate}

\noindent
\textbf{Threats to Internal Validity} mainly contains in two-folds.
The first one is the design of a prompt to instruct ChatGPT to give out responses.
We design our prompt according to the practical advice~\cite{shieh2023best} which has been verified by many users online and can obtain a good response from ChatGPT.
Furthermore, ChatGPT will generate responses with some randomness even given the same prompt.
Therefore, we set ``temperature'' to 0, which will reduce the randomness at most and we try our best to collect all these results in two days to avoid the model being upgraded. 
The second one is about the potential mistakes in the implementation of studied baselines. 
To minimize such threats, we directly use the original source code shared by corresponding authors.

\noindent
\textbf{Threats to External Validity} may correspond to the generalization of the studied dataset.
To mitigate this threat, we adopt the most large-scale vulnerability dataset with diverse information about the vulnerabilities, which are collected from practical projects, and these vulnerabilities are recorded in the Common Vulnerabilities and Exposures (CVE).
However, we do not consider these vulnerabilities found recently.
Besides, we do not adopt another large-scale vulnerability dataset named SARD since it is built manually and cannot satisfy the distinct characteristics of the real world~\cite{hin2022linevd,chakraborty2021deep}.

\noindent
\textbf{Threats to Construct Validity} mainly correspond to  the performance metrics in our evaluations.
To minimize such threats, we consider a few widely used performance metrics to evaluate the performance of ChatGPT on different types of tasks. (e.g., Accuracy, Precision, Recall, and ROUGE).

%% file: sections/conclusion.tex
\section{Conclusion}

This paper aims to comprehensively investigate the capabilities of ChatGPT for software vulnerability tasks as well as its impacts.
To achieve that, we adopt a large-scale vulnerability dataset (named Big-Vul) and then conduct several experiments focusing on five dimensions: 
\textbf{(1) Vulnerability Detection}, \textbf{(2) Vulnerability Assessment}, \textbf{(3) Vulnerability Localization}, \textbf{(4) Vulnerability Repair}, and \textbf{(5) Vulnerability Description}.
Overall, although ChatGPT shows some ability in certain areas, it still needs further improvement to be competent in software vulnerability related tasks.
Our research conducts a comprehensive survey of ChatGPT's capabilities and provides a reference for enhancing its understanding of software vulnerabilities in the future.

%% file: main.bbl

\begin{thebibliography}{61}


\ifx \showCODEN    \undefined \def \showCODEN     #1{\unskip}     \fi
\ifx \showDOI      \undefined \def \showDOI       #1{#1}\fi
\ifx \showISBNx    \undefined \def \showISBNx     #1{\unskip}     \fi
\ifx \showISBNxiii \undefined \def \showISBNxiii  #1{\unskip}     \fi
\ifx \showISSN     \undefined \def \showISSN      #1{\unskip}     \fi
\ifx \showLCCN     \undefined \def \showLCCN      #1{\unskip}     \fi
\ifx \shownote     \undefined \def \shownote      #1{#1}          \fi
\ifx \showarticletitle \undefined \def \showarticletitle #1{#1}   \fi
\ifx \showURL      \undefined \def \showURL       {\relax}        \fi
\providecommand\bibfield[2]{#2}
\providecommand\bibinfo[2]{#2}
\providecommand\natexlab[1]{#1}
\providecommand\showeprint[2][]{arXiv:#2}

\bibitem[joe(2023)]%
        {joern}
 \bibinfo{year}{2023}\natexlab{}.
\newblock \bibinfo{title}{Joern}.
\newblock
\newblock
\urldef\tempurl%
\url{https://github.com/joernio/joern}
\showURL{%
\tempurl}


\bibitem[rep(2023)]%
        {replication}
 \bibinfo{year}{2023}\natexlab{}.
\newblock \bibinfo{title}{Replication}.
\newblock
\newblock
\urldef\tempurl%
\url{https://figshare.com/s/04856ae0c9005a888e03}
\showURL{%
\tempurl}


\bibitem[Ahmad et~al\mbox{.}(2021)]%
        {ahmad2021unified}
\bibfield{author}{\bibinfo{person}{Wasi~Uddin Ahmad}, \bibinfo{person}{Saikat Chakraborty}, \bibinfo{person}{Baishakhi Ray}, {and} \bibinfo{person}{Kai-Wei Chang}.} \bibinfo{year}{2021}\natexlab{}.
\newblock \showarticletitle{Unified pre-training for program understanding and generation}.
\newblock \bibinfo{journal}{\emph{arXiv preprint arXiv:2103.06333}} (\bibinfo{year}{2021}).
\newblock


\bibitem[Bang et~al\mbox{.}(2023)]%
        {bang2023multitask}
\bibfield{author}{\bibinfo{person}{Yejin Bang}, \bibinfo{person}{Samuel Cahyawijaya}, \bibinfo{person}{Nayeon Lee}, \bibinfo{person}{Wenliang Dai}, \bibinfo{person}{Dan Su}, \bibinfo{person}{Bryan Wilie}, \bibinfo{person}{Holy Lovenia}, \bibinfo{person}{Ziwei Ji}, \bibinfo{person}{Tiezheng Yu}, \bibinfo{person}{Willy Chung}, {et~al\mbox{.}}} \bibinfo{year}{2023}\natexlab{}.
\newblock \showarticletitle{A multitask, multilingual, multimodal evaluation of chatgpt on reasoning, hallucination, and interactivity}.
\newblock \bibinfo{journal}{\emph{arXiv preprint arXiv:2302.04023}} (\bibinfo{year}{2023}).
\newblock


\bibitem[Bhandari et~al\mbox{.}(2021)]%
        {bhandari2021cvefixes}
\bibfield{author}{\bibinfo{person}{Guru Bhandari}, \bibinfo{person}{Amara Naseer}, {and} \bibinfo{person}{Leon Moonen}.} \bibinfo{year}{2021}\natexlab{}.
\newblock \showarticletitle{CVEfixes: automated collection of vulnerabilities and their fixes from open-source software}. In \bibinfo{booktitle}{\emph{Proceedings of the 17th International Conference on Predictive Models and Data Analytics in Software Engineering}}. \bibinfo{pages}{30--39}.
\newblock


\bibitem[Brown et~al\mbox{.}(2020)]%
        {brown2020language}
\bibfield{author}{\bibinfo{person}{Tom Brown}, \bibinfo{person}{Benjamin Mann}, \bibinfo{person}{Nick Ryder}, \bibinfo{person}{Melanie Subbiah}, \bibinfo{person}{Jared~D Kaplan}, \bibinfo{person}{Prafulla Dhariwal}, \bibinfo{person}{Arvind Neelakantan}, \bibinfo{person}{Pranav Shyam}, \bibinfo{person}{Girish Sastry}, \bibinfo{person}{Amanda Askell}, {et~al\mbox{.}}} \bibinfo{year}{2020}\natexlab{}.
\newblock \showarticletitle{Language models are few-shot learners}.
\newblock \bibinfo{journal}{\emph{Advances in neural information processing systems}}  \bibinfo{volume}{33} (\bibinfo{year}{2020}), \bibinfo{pages}{1877--1901}.
\newblock


\bibitem[Cao et~al\mbox{.}(2022)]%
        {cao2022mvd}
\bibfield{author}{\bibinfo{person}{Sicong Cao}, \bibinfo{person}{Xiaobing Sun}, \bibinfo{person}{Lili Bo}, \bibinfo{person}{Rongxin Wu}, \bibinfo{person}{Bin Li}, {and} \bibinfo{person}{Chuanqi Tao}.} \bibinfo{year}{2022}\natexlab{}.
\newblock \showarticletitle{MVD: Memory-Related Vulnerability Detection Based on Flow-Sensitive Graph Neural Networks}.
\newblock \bibinfo{journal}{\emph{arXiv preprint arXiv:2203.02660}} (\bibinfo{year}{2022}).
\newblock


\bibitem[Chakraborty et~al\mbox{.}(2021)]%
        {chakraborty2021deep}
\bibfield{author}{\bibinfo{person}{Saikat Chakraborty}, \bibinfo{person}{Rahul Krishna}, \bibinfo{person}{Yangruibo Ding}, {and} \bibinfo{person}{Baishakhi Ray}.} \bibinfo{year}{2021}\natexlab{}.
\newblock \showarticletitle{Deep learning based vulnerability detection: Are we there yet}.
\newblock \bibinfo{journal}{\emph{IEEE Transactions on Software Engineering}} (\bibinfo{year}{2021}).
\newblock


\bibitem[chatgptendpoint(2023)]%
        {2023chatgptendpoint}
\bibfield{author}{\bibinfo{person}{chatgptendpoint}.} \bibinfo{year}{2023}\natexlab{}.
\newblock \bibinfo{title}{Introducing ChatGPT and Whisper APIs}.
\newblock \bibinfo{howpublished}{\url{https://openai.com/blog/introducing-chatgpt-and-whisper-apis}}.
\newblock


\bibitem[Chen et~al\mbox{.}(2019)]%
        {chen2019sequencer}
\bibfield{author}{\bibinfo{person}{Zimin Chen}, \bibinfo{person}{Steve~James Kommrusch}, \bibinfo{person}{Michele Tufano}, \bibinfo{person}{Louis-No{\"e}l Pouchet}, \bibinfo{person}{Denys Poshyvanyk}, {and} \bibinfo{person}{Martin Monperrus}.} \bibinfo{year}{2019}\natexlab{}.
\newblock \showarticletitle{Sequencer: Sequence-to-sequence learning for end-to-end program repair}.
\newblock \bibinfo{journal}{\emph{IEEE Transactions on Software Engineering}} (\bibinfo{year}{2019}).
\newblock


\bibitem[Cheng et~al\mbox{.}(2022)]%
        {cheng2022path}
\bibfield{author}{\bibinfo{person}{Xiao Cheng}, \bibinfo{person}{Guanqin Zhang}, \bibinfo{person}{Haoyu Wang}, {and} \bibinfo{person}{Yulei Sui}.} \bibinfo{year}{2022}\natexlab{}.
\newblock \showarticletitle{Path-sensitive code embedding via contrastive learning for software vulnerability detection}. In \bibinfo{booktitle}{\emph{Proceedings of the 31st ACM SIGSOFT International Symposium on Software Testing and Analysis}}. \bibinfo{pages}{519--531}.
\newblock


\bibitem[Choi et~al\mbox{.}(2023)]%
        {choi2023chatgpt}
\bibfield{author}{\bibinfo{person}{Jonathan~H Choi}, \bibinfo{person}{Kristin~E Hickman}, \bibinfo{person}{Amy Monahan}, {and} \bibinfo{person}{Daniel Schwarcz}.} \bibinfo{year}{2023}\natexlab{}.
\newblock \showarticletitle{Chatgpt goes to law school}.
\newblock \bibinfo{journal}{\emph{Available at SSRN}} (\bibinfo{year}{2023}).
\newblock


\bibitem[Christiano et~al\mbox{.}(2017)]%
        {christiano2017deep}
\bibfield{author}{\bibinfo{person}{Paul~F Christiano}, \bibinfo{person}{Jan Leike}, \bibinfo{person}{Tom Brown}, \bibinfo{person}{Miljan Martic}, \bibinfo{person}{Shane Legg}, {and} \bibinfo{person}{Dario Amodei}.} \bibinfo{year}{2017}\natexlab{}.
\newblock \showarticletitle{Deep reinforcement learning from human preferences}.
\newblock \bibinfo{journal}{\emph{Advances in neural information processing systems}}  \bibinfo{volume}{30} (\bibinfo{year}{2017}).
\newblock


\bibitem[Fan et~al\mbox{.}(2019)]%
        {fan2019smoke}
\bibfield{author}{\bibinfo{person}{Gang Fan}, \bibinfo{person}{Rongxin Wu}, \bibinfo{person}{Qingkai Shi}, \bibinfo{person}{Xiao Xiao}, \bibinfo{person}{Jinguo Zhou}, {and} \bibinfo{person}{Charles Zhang}.} \bibinfo{year}{2019}\natexlab{}.
\newblock \showarticletitle{Smoke: scalable path-sensitive memory leak detection for millions of lines of code}. In \bibinfo{booktitle}{\emph{2019 IEEE/ACM 41st International Conference on Software Engineering (ICSE)}}. IEEE, \bibinfo{pages}{72--82}.
\newblock


\bibitem[Fan et~al\mbox{.}(2020)]%
        {fan2020ac}
\bibfield{author}{\bibinfo{person}{Jiahao Fan}, \bibinfo{person}{Yi Li}, \bibinfo{person}{Shaohua Wang}, {and} \bibinfo{person}{Tien~N Nguyen}.} \bibinfo{year}{2020}\natexlab{}.
\newblock \showarticletitle{A C/C++ code vulnerability dataset with code changes and CVE summaries}. In \bibinfo{booktitle}{\emph{Proceedings of the 17th International Conference on Mining Software Repositories}}. \bibinfo{pages}{508--512}.
\newblock


\bibitem[Feng et~al\mbox{.}(2020)]%
        {feng2020codebert}
\bibfield{author}{\bibinfo{person}{Zhangyin Feng}, \bibinfo{person}{Daya Guo}, \bibinfo{person}{Duyu Tang}, \bibinfo{person}{Nan Duan}, \bibinfo{person}{Xiaocheng Feng}, \bibinfo{person}{Ming Gong}, \bibinfo{person}{Linjun Shou}, \bibinfo{person}{Bing Qin}, \bibinfo{person}{Ting Liu}, \bibinfo{person}{Daxin Jiang}, {et~al\mbox{.}}} \bibinfo{year}{2020}\natexlab{}.
\newblock \showarticletitle{Codebert: A pre-trained model for programming and natural languages}.
\newblock \bibinfo{journal}{\emph{arXiv preprint arXiv:2002.08155}} (\bibinfo{year}{2020}).
\newblock


\bibitem[Feutrill et~al\mbox{.}(2018)]%
        {feutrill2018effect}
\bibfield{author}{\bibinfo{person}{Andrew Feutrill}, \bibinfo{person}{Dinesha Ranathunga}, \bibinfo{person}{Yuval Yarom}, {and} \bibinfo{person}{Matthew Roughan}.} \bibinfo{year}{2018}\natexlab{}.
\newblock \showarticletitle{The effect of common vulnerability scoring system metrics on vulnerability exploit delay}. In \bibinfo{booktitle}{\emph{2018 Sixth International Symposium on Computing and Networking (CANDAR)}}. IEEE, \bibinfo{pages}{1--10}.
\newblock


\bibitem[Fu and Tantithamthavorn(2022)]%
        {fu2022linevul}
\bibfield{author}{\bibinfo{person}{Michael Fu} {and} \bibinfo{person}{Chakkrit Tantithamthavorn}.} \bibinfo{year}{2022}\natexlab{}.
\newblock \showarticletitle{LineVul: A Transformer-based Line-Level Vulnerability Prediction}.
\newblock  (\bibinfo{year}{2022}).
\newblock


\bibitem[Gilson et~al\mbox{.}(2022)]%
        {gilson2022well}
\bibfield{author}{\bibinfo{person}{Aidan Gilson}, \bibinfo{person}{Conrad Safranek}, \bibinfo{person}{Thomas Huang}, \bibinfo{person}{Vimig Socrates}, \bibinfo{person}{Ling Chi}, \bibinfo{person}{Richard~Andrew Taylor}, {and} \bibinfo{person}{David Chartash}.} \bibinfo{year}{2022}\natexlab{}.
\newblock \showarticletitle{How Well Does ChatGPT Do When Taking the Medical Licensing Exams? The Implications of Large Language Models for Medical Education and Knowledge Assessment}.
\newblock \bibinfo{journal}{\emph{medRxiv}} (\bibinfo{year}{2022}), \bibinfo{pages}{2022--12}.
\newblock


\bibitem[Goldberg(2023a)]%
        {firendorfoe}
\bibfield{author}{\bibinfo{person}{Yoav Goldberg}.} \bibinfo{year}{2023}\natexlab{a}.
\newblock \bibinfo{title}{Friend or foe? teachers debate chatgpt.}
\newblock
\newblock
\urldef\tempurl%
\url{https://www.axios.com/2023/01/13/chatgpt-schools-teachers-ai-debate}
\showURL{%
\tempurl}


\bibitem[Goldberg(2023b)]%
        {remarkonllm}
\bibfield{author}{\bibinfo{person}{Yoav Goldberg}.} \bibinfo{year}{2023}\natexlab{b}.
\newblock \bibinfo{title}{Some remarks on large language models}.
\newblock
\newblock
\urldef\tempurl%
\url{https://gist.github.com/yoavg/59d174608e92e845c8994ac2e234c8a9}
\showURL{%
\tempurl}


\bibitem[Guo et~al\mbox{.}(2022b)]%
        {guo2022unixcoder}
\bibfield{author}{\bibinfo{person}{Daya Guo}, \bibinfo{person}{Shuai Lu}, \bibinfo{person}{Nan Duan}, \bibinfo{person}{Yanlin Wang}, \bibinfo{person}{Ming Zhou}, {and} \bibinfo{person}{Jian Yin}.} \bibinfo{year}{2022}\natexlab{b}.
\newblock \showarticletitle{UniXcoder: Unified Cross-Modal Pre-training for Code Representation}.
\newblock \bibinfo{journal}{\emph{arXiv preprint arXiv:2203.03850}} (\bibinfo{year}{2022}).
\newblock


\bibitem[Guo et~al\mbox{.}(2020a)]%
        {guo2020graphcodebert}
\bibfield{author}{\bibinfo{person}{Daya Guo}, \bibinfo{person}{Shuo Ren}, \bibinfo{person}{Shuai Lu}, \bibinfo{person}{Zhangyin Feng}, \bibinfo{person}{Duyu Tang}, \bibinfo{person}{Shujie Liu}, \bibinfo{person}{Long Zhou}, \bibinfo{person}{Nan Duan}, \bibinfo{person}{Alexey Svyatkovskiy}, \bibinfo{person}{Shengyu Fu}, {et~al\mbox{.}}} \bibinfo{year}{2020}\natexlab{a}.
\newblock \showarticletitle{Graphcodebert: Pre-training code representations with data flow}.
\newblock \bibinfo{journal}{\emph{arXiv preprint arXiv:2009.08366}} (\bibinfo{year}{2020}).
\newblock


\bibitem[Guo et~al\mbox{.}(2022a)]%
        {guo2022detecting}
\bibfield{author}{\bibinfo{person}{Hao Guo}, \bibinfo{person}{Sen Chen}, \bibinfo{person}{Zhenchang Xing}, \bibinfo{person}{Xiaohong Li}, \bibinfo{person}{Yude Bai}, {and} \bibinfo{person}{Jiamou Sun}.} \bibinfo{year}{2022}\natexlab{a}.
\newblock \showarticletitle{Detecting and augmenting missing key aspects in vulnerability descriptions}.
\newblock \bibinfo{journal}{\emph{ACM Transactions on Software Engineering and Methodology (TOSEM)}} \bibinfo{volume}{31}, \bibinfo{number}{3} (\bibinfo{year}{2022}), \bibinfo{pages}{1--27}.
\newblock


\bibitem[Guo et~al\mbox{.}(2021)]%
        {guo2021key}
\bibfield{author}{\bibinfo{person}{Hao Guo}, \bibinfo{person}{Zhenchang Xing}, \bibinfo{person}{Sen Chen}, \bibinfo{person}{Xiaohong Li}, \bibinfo{person}{Yude Bai}, {and} \bibinfo{person}{Hu Zhang}.} \bibinfo{year}{2021}\natexlab{}.
\newblock \showarticletitle{Key aspects augmentation of vulnerability description based on multiple security databases}. In \bibinfo{booktitle}{\emph{2021 IEEE 45th Annual Computers, Software, and Applications Conference (COMPSAC)}}. IEEE, \bibinfo{pages}{1020--1025}.
\newblock


\bibitem[Guo et~al\mbox{.}(2020b)]%
        {guo2020predicting}
\bibfield{author}{\bibinfo{person}{Hao Guo}, \bibinfo{person}{Zhenchang Xing}, {and} \bibinfo{person}{Xiaohong Li}.} \bibinfo{year}{2020}\natexlab{b}.
\newblock \showarticletitle{Predicting missing information of key aspects in vulnerability reports}.
\newblock \bibinfo{journal}{\emph{arXiv preprint arXiv:2008.02456}} (\bibinfo{year}{2020}).
\newblock


\bibitem[Hin et~al\mbox{.}(2022)]%
        {hin2022linevd}
\bibfield{author}{\bibinfo{person}{David Hin}, \bibinfo{person}{Andrey Kan}, \bibinfo{person}{Huaming Chen}, {and} \bibinfo{person}{M~Ali Babar}.} \bibinfo{year}{2022}\natexlab{}.
\newblock \showarticletitle{LineVD: Statement-level Vulnerability Detection using Graph Neural Networks}.
\newblock \bibinfo{journal}{\emph{arXiv preprint arXiv:2203.05181}} (\bibinfo{year}{2022}).
\newblock


\bibitem[Khan and Parkinson(2018)]%
        {khan2018review}
\bibfield{author}{\bibinfo{person}{Saad Khan} {and} \bibinfo{person}{Simon Parkinson}.} \bibinfo{year}{2018}\natexlab{}.
\newblock \showarticletitle{Review into state of the art of vulnerability assessment using artificial intelligence}.
\newblock \bibinfo{journal}{\emph{Guide to Vulnerability Analysis for Computer Networks and Systems}} (\bibinfo{year}{2018}), \bibinfo{pages}{3--32}.
\newblock


\bibitem[Le et~al\mbox{.}(2021a)]%
        {le2021survey}
\bibfield{author}{\bibinfo{person}{Triet~HM Le}, \bibinfo{person}{Huaming Chen}, {and} \bibinfo{person}{M~Ali Babar}.} \bibinfo{year}{2021}\natexlab{a}.
\newblock \showarticletitle{A survey on data-driven software vulnerability assessment and prioritization}.
\newblock \bibinfo{journal}{\emph{ACM Computing Surveys (CSUR)}} (\bibinfo{year}{2021}).
\newblock


\bibitem[Le et~al\mbox{.}(2021b)]%
        {le2021deepcva}
\bibfield{author}{\bibinfo{person}{Triet Huynh~Minh Le}, \bibinfo{person}{David Hin}, \bibinfo{person}{Roland Croft}, {and} \bibinfo{person}{M~Ali Babar}.} \bibinfo{year}{2021}\natexlab{b}.
\newblock \showarticletitle{Deepcva: Automated commit-level vulnerability assessment with deep multi-task learning}. In \bibinfo{booktitle}{\emph{2021 36th IEEE/ACM International Conference on Automated Software Engineering (ASE)}}. IEEE, \bibinfo{pages}{717--729}.
\newblock


\bibitem[Li et~al\mbox{.}(2020)]%
        {li2020pca}
\bibfield{author}{\bibinfo{person}{Wen Li}, \bibinfo{person}{Haipeng Cai}, \bibinfo{person}{Yulei Sui}, {and} \bibinfo{person}{David Manz}.} \bibinfo{year}{2020}\natexlab{}.
\newblock \showarticletitle{PCA: memory leak detection using partial call-path analysis}. In \bibinfo{booktitle}{\emph{Proceedings of the 28th ACM Joint Meeting on European Software Engineering Conference and Symposium on the Foundations of Software Engineering}}. \bibinfo{pages}{1621--1625}.
\newblock


\bibitem[Li et~al\mbox{.}(2021a)]%
        {li2021vulnerability}
\bibfield{author}{\bibinfo{person}{Yi Li}, \bibinfo{person}{Shaohua Wang}, {and} \bibinfo{person}{Tien~N Nguyen}.} \bibinfo{year}{2021}\natexlab{a}.
\newblock \showarticletitle{Vulnerability detection with fine-grained interpretations}. In \bibinfo{booktitle}{\emph{Proceedings of the 29th ACM Joint Meeting on European Software Engineering Conference and Symposium on the Foundations of Software Engineering}}. \bibinfo{pages}{292--303}.
\newblock


\bibitem[Li et~al\mbox{.}(2021b)]%
        {li2021vuldeelocator}
\bibfield{author}{\bibinfo{person}{Zhen Li}, \bibinfo{person}{Deqing Zou}, \bibinfo{person}{Shouhuai Xu}, \bibinfo{person}{Zhaoxuan Chen}, \bibinfo{person}{Yawei Zhu}, {and} \bibinfo{person}{Hai Jin}.} \bibinfo{year}{2021}\natexlab{b}.
\newblock \showarticletitle{Vuldeelocator: a deep learning-based fine-grained vulnerability detector}.
\newblock \bibinfo{journal}{\emph{IEEE Transactions on Dependable and Secure Computing}} (\bibinfo{year}{2021}).
\newblock


\bibitem[Li et~al\mbox{.}(2021c)]%
        {li2021sysevr}
\bibfield{author}{\bibinfo{person}{Zhen Li}, \bibinfo{person}{Deqing Zou}, \bibinfo{person}{Shouhuai Xu}, \bibinfo{person}{Hai Jin}, \bibinfo{person}{Yawei Zhu}, {and} \bibinfo{person}{Zhaoxuan Chen}.} \bibinfo{year}{2021}\natexlab{c}.
\newblock \showarticletitle{Sysevr: A framework for using deep learning to detect software vulnerabilities}.
\newblock \bibinfo{journal}{\emph{IEEE Transactions on Dependable and Secure Computing}} (\bibinfo{year}{2021}).
\newblock


\bibitem[Li et~al\mbox{.}(2018)]%
        {li2018vuldeepecker}
\bibfield{author}{\bibinfo{person}{Zhen Li}, \bibinfo{person}{Deqing Zou}, \bibinfo{person}{Shouhuai Xu}, \bibinfo{person}{Xinyu Ou}, \bibinfo{person}{Hai Jin}, \bibinfo{person}{Sujuan Wang}, \bibinfo{person}{Zhijun Deng}, {and} \bibinfo{person}{Yuyi Zhong}.} \bibinfo{year}{2018}\natexlab{}.
\newblock \showarticletitle{Vuldeepecker: A deep learning-based system for vulnerability detection}. In \bibinfo{booktitle}{\emph{Proceedings of the 25th Annual Network and Distributed System Security Symposium}}.
\newblock


\bibitem[Lin(2004)]%
        {lin2004rouge}
\bibfield{author}{\bibinfo{person}{Chin-Yew Lin}.} \bibinfo{year}{2004}\natexlab{}.
\newblock \showarticletitle{Rouge: A package for automatic evaluation of summaries}. In \bibinfo{booktitle}{\emph{Text summarization branches out}}. \bibinfo{pages}{74--81}.
\newblock


\bibitem[Liu et~al\mbox{.}(2023)]%
        {liu2023pre}
\bibfield{author}{\bibinfo{person}{Pengfei Liu}, \bibinfo{person}{Weizhe Yuan}, \bibinfo{person}{Jinlan Fu}, \bibinfo{person}{Zhengbao Jiang}, \bibinfo{person}{Hiroaki Hayashi}, {and} \bibinfo{person}{Graham Neubig}.} \bibinfo{year}{2023}\natexlab{}.
\newblock \showarticletitle{Pre-train, prompt, and predict: A systematic survey of prompting methods in natural language processing}.
\newblock \bibinfo{journal}{\emph{Comput. Surveys}} \bibinfo{volume}{55}, \bibinfo{number}{9} (\bibinfo{year}{2023}), \bibinfo{pages}{1--35}.
\newblock


\bibitem[MITRE(2023)]%
        {cve}
\bibfield{author}{\bibinfo{person}{Corporation MITRE}.} \bibinfo{year}{2023}\natexlab{}.
\newblock \bibinfo{title}{Common Vulnerabilities and Exposures (CVE)}.
\newblock
\newblock
\urldef\tempurl%
\url{https://cve.mitre.org/}
\showURL{%
\tempurl}


\bibitem[Ni et~al\mbox{.}(2023a)]%
        {ni2023fva}
\bibfield{author}{\bibinfo{person}{Chao Ni}, \bibinfo{person}{Liyu Shen}, \bibinfo{person}{Wei Wang}, \bibinfo{person}{Xiang Chen}, \bibinfo{person}{Xin Yin}, {and} \bibinfo{person}{Lexiao Zhang}.} \bibinfo{year}{2023}\natexlab{a}.
\newblock \showarticletitle{FVA: Assessing Function-Level Vulnerability by Integrating Flow-Sensitive Structure and Code Statement Semantic}. In \bibinfo{booktitle}{\emph{2023 IEEE/ACM 31st International Conference on Program Comprehension (ICPC)}}. IEEE, \bibinfo{pages}{339--350}.
\newblock


\bibitem[Ni et~al\mbox{.}(2022a)]%
        {ni2022best}
\bibfield{author}{\bibinfo{person}{Chao Ni}, \bibinfo{person}{Wei Wang}, \bibinfo{person}{Kaiwen Yang}, \bibinfo{person}{Xin Xia}, \bibinfo{person}{Kui Liu}, {and} \bibinfo{person}{David Lo}.} \bibinfo{year}{2022}\natexlab{a}.
\newblock \showarticletitle{{ The Best of Both Worlds: Integrating Semantic Features with Expert Features for Defect Prediction and Localization}}. In \bibinfo{booktitle}{\emph{Proceedings of the 2022 30th ACM Joint Meeting on European Software Engineering Conference and Symposium on the Foundations of Software Engineering}}. ACM, \bibinfo{pages}{672--683}.
\newblock


\bibitem[Ni et~al\mbox{.}(2022b)]%
        {ni2022defect}
\bibfield{author}{\bibinfo{person}{Chao Ni}, \bibinfo{person}{Kaiwen Yang}, \bibinfo{person}{Xin Xia}, \bibinfo{person}{David Lo}, \bibinfo{person}{Xiang Chen}, {and} \bibinfo{person}{Xiaohu Yang}.} \bibinfo{year}{2022}\natexlab{b}.
\newblock \showarticletitle{Defect Identification, Categorization, and Repair: Better Together}.
\newblock \bibinfo{journal}{\emph{arXiv preprint arXiv:2204.04856}} (\bibinfo{year}{2022}).
\newblock


\bibitem[Ni et~al\mbox{.}(2023b)]%
        {ni2023distinguishing}
\bibfield{author}{\bibinfo{person}{Chao Ni}, \bibinfo{person}{Xin Yin}, \bibinfo{person}{Kaiwen Yang}, \bibinfo{person}{Dehai Zhao}, \bibinfo{person}{Zhenchang Xing}, {and} \bibinfo{person}{Xin Xia}.} \bibinfo{year}{2023}\natexlab{b}.
\newblock \showarticletitle{Distinguishing Look-Alike Innocent and Vulnerable Code by Subtle Semantic Representation Learning and Explanation}. In \bibinfo{booktitle}{\emph{Proceedings of the 31st ACM Joint European Software Engineering Conference and Symposium on the Foundations of Software Engineering}}. \bibinfo{pages}{1611--1622}.
\newblock


\bibitem[OpenAI(2022)]%
        {openai2022chatgpt}
\bibfield{author}{\bibinfo{person}{OpenAI}.} \bibinfo{year}{2022}\natexlab{}.
\newblock \bibinfo{title}{ChatGPT: Optimizing Language Models for Dialogue. (2022)}.
\newblock \bibinfo{howpublished}{\url{https://openai.com/blog/chatgpt/}}.
\newblock


\bibitem[Ouyang et~al\mbox{.}(2022)]%
        {ouyang2022training}
\bibfield{author}{\bibinfo{person}{Long Ouyang}, \bibinfo{person}{Jeffrey Wu}, \bibinfo{person}{Xu Jiang}, \bibinfo{person}{Diogo Almeida}, \bibinfo{person}{Carroll Wainwright}, \bibinfo{person}{Pamela Mishkin}, \bibinfo{person}{Chong Zhang}, \bibinfo{person}{Sandhini Agarwal}, \bibinfo{person}{Katarina Slama}, \bibinfo{person}{Alex Ray}, {et~al\mbox{.}}} \bibinfo{year}{2022}\natexlab{}.
\newblock \showarticletitle{Training language models to follow instructions with human feedback}.
\newblock \bibinfo{journal}{\emph{Advances in Neural Information Processing Systems}}  \bibinfo{volume}{35} (\bibinfo{year}{2022}), \bibinfo{pages}{27730--27744}.
\newblock


\bibitem[Paszke et~al\mbox{.}(2019)]%
        {pytorch}
\bibfield{author}{\bibinfo{person}{Adam Paszke}, \bibinfo{person}{Sam Gross}, \bibinfo{person}{Francisco Massa}, \bibinfo{person}{Adam Lerer}, \bibinfo{person}{James Bradbury}, \bibinfo{person}{Gregory Chanan}, \bibinfo{person}{Trevor Killeen}, \bibinfo{person}{Zeming Lin}, \bibinfo{person}{Natalia Gimelshein}, \bibinfo{person}{Luca Antiga}, \bibinfo{person}{Alban Desmaison}, \bibinfo{person}{Andreas Kopf}, \bibinfo{person}{Edward Yang}, \bibinfo{person}{Zachary DeVito}, \bibinfo{person}{Martin Raison}, \bibinfo{person}{Alykhan Tejani}, \bibinfo{person}{Sasank Chilamkurthy}, \bibinfo{person}{Benoit Steiner}, \bibinfo{person}{Lu Fang}, \bibinfo{person}{Junjie Bai}, {and} \bibinfo{person}{Soumith Chintala}.} \bibinfo{year}{2019}\natexlab{}.
\newblock \showarticletitle{PyTorch: An Imperative Style, High-Performance Deep Learning Library}.
\newblock In \bibinfo{booktitle}{\emph{Advances in Neural Information Processing Systems 32}}. \bibinfo{publisher}{Curran Associates, Inc.}, \bibinfo{pages}{8024--8035}.
\newblock
\urldef\tempurl%
\url{http://papers.neurips.cc/paper/9015-pytorch-an-imperative-style-high-performance-deep-learning-library.pdf}
\showURL{%
\tempurl}


\bibitem[Radford et~al\mbox{.}(2018)]%
        {radford2018improving}
\bibfield{author}{\bibinfo{person}{Alec Radford}, \bibinfo{person}{Karthik Narasimhan}, \bibinfo{person}{Tim Salimans}, \bibinfo{person}{Ilya Sutskever}, {et~al\mbox{.}}} \bibinfo{year}{2018}\natexlab{}.
\newblock \showarticletitle{Improving language understanding by generative pre-training}.
\newblock  (\bibinfo{year}{2018}).
\newblock


\bibitem[Schulman et~al\mbox{.}(2017)]%
        {schulman2017proximal}
\bibfield{author}{\bibinfo{person}{John Schulman}, \bibinfo{person}{Filip Wolski}, \bibinfo{person}{Prafulla Dhariwal}, \bibinfo{person}{Alec Radford}, {and} \bibinfo{person}{Oleg Klimov}.} \bibinfo{year}{2017}\natexlab{}.
\newblock \showarticletitle{Proximal policy optimization algorithms}.
\newblock \bibinfo{journal}{\emph{arXiv preprint arXiv:1707.06347}} (\bibinfo{year}{2017}).
\newblock


\bibitem[Shen et~al\mbox{.}(2023)]%
        {shen2023chatgpt}
\bibfield{author}{\bibinfo{person}{Yiqiu Shen}, \bibinfo{person}{Laura Heacock}, \bibinfo{person}{Jonathan Elias}, \bibinfo{person}{Keith~D Hentel}, \bibinfo{person}{Beatriu Reig}, \bibinfo{person}{George Shih}, {and} \bibinfo{person}{Linda Moy}.} \bibinfo{year}{2023}\natexlab{}.
\newblock \bibinfo{title}{ChatGPT and other large language models are double-edged swords}.
\newblock , \bibinfo{numpages}{230163}~pages.
\newblock


\bibitem[Shieh(2023)]%
        {shieh2023best}
\bibfield{author}{\bibinfo{person}{Jessica Shieh}.} \bibinfo{year}{2023}\natexlab{}.
\newblock \showarticletitle{Best practices for prompt engineering with OpenAI API}.
\newblock \bibinfo{journal}{\emph{OpenAI, February https://help.openai. com/en/articles/6654000-best-practices-for-prompt-engineering-with-openai-api}} (\bibinfo{year}{2023}).
\newblock


\bibitem[Spanos and Angelis(2018)]%
        {spanos2018multi}
\bibfield{author}{\bibinfo{person}{Georgios Spanos} {and} \bibinfo{person}{Lefteris Angelis}.} \bibinfo{year}{2018}\natexlab{}.
\newblock \showarticletitle{A multi-target approach to estimate software vulnerability characteristics and severity scores}.
\newblock \bibinfo{journal}{\emph{Journal of Systems and Software}}  \bibinfo{volume}{146} (\bibinfo{year}{2018}), \bibinfo{pages}{152--166}.
\newblock


\bibitem[Sun et~al\mbox{.}(2022)]%
        {sun2021generating}
\bibfield{author}{\bibinfo{person}{Jiamou Sun}, \bibinfo{person}{Zhenchang Xing}, \bibinfo{person}{Hao Guo}, \bibinfo{person}{Deheng Ye}, \bibinfo{person}{Xiaohong Li}, \bibinfo{person}{Xiwei Xu}, {and} \bibinfo{person}{Liming Zhu}.} \bibinfo{year}{2022}\natexlab{}.
\newblock \showarticletitle{Generating informative CVE description from ExploitDB posts by extractive summarization}.
\newblock \bibinfo{journal}{\emph{ACM Transactions on Software Engineering and Methodology (TOSEM)}} (\bibinfo{year}{2022}).
\newblock


\bibitem[Symantec(2023)]%
        {securityfocus}
\bibfield{author}{\bibinfo{person}{Symantec}.} \bibinfo{year}{2023}\natexlab{}.
\newblock \bibinfo{title}{securityFocus}.
\newblock
\newblock
\urldef\tempurl%
\url{https://www.securityfocus.com/}
\showURL{%
\tempurl}


\bibitem[Vaswani et~al\mbox{.}(2017)]%
        {vaswani2017attention}
\bibfield{author}{\bibinfo{person}{Ashish Vaswani}, \bibinfo{person}{Noam Shazeer}, \bibinfo{person}{Niki Parmar}, \bibinfo{person}{Jakob Uszkoreit}, \bibinfo{person}{Llion Jones}, \bibinfo{person}{Aidan~N Gomez}, \bibinfo{person}{{\L}ukasz Kaiser}, {and} \bibinfo{person}{Illia Polosukhin}.} \bibinfo{year}{2017}\natexlab{}.
\newblock \showarticletitle{Attention is all you need}.
\newblock \bibinfo{journal}{\emph{Advances in neural information processing systems}}  \bibinfo{volume}{30} (\bibinfo{year}{2017}).
\newblock


\bibitem[Wang et~al\mbox{.}(2021)]%
        {wang2021codet5}
\bibfield{author}{\bibinfo{person}{Yue Wang}, \bibinfo{person}{Weishi Wang}, \bibinfo{person}{Shafiq Joty}, {and} \bibinfo{person}{Steven~CH Hoi}.} \bibinfo{year}{2021}\natexlab{}.
\newblock \showarticletitle{Codet5: Identifier-aware unified pre-trained encoder-decoder models for code understanding and generation}.
\newblock \bibinfo{journal}{\emph{arXiv preprint arXiv:2109.00859}} (\bibinfo{year}{2021}).
\newblock


\bibitem[Wu et~al\mbox{.}(2022)]%
        {wu2022vulcnn}
\bibfield{author}{\bibinfo{person}{Yueming Wu}, \bibinfo{person}{Deqing Zou}, \bibinfo{person}{Shihan Dou}, \bibinfo{person}{Wei Yang}, \bibinfo{person}{Duo Xu}, {and} \bibinfo{person}{Hai Jin}.} \bibinfo{year}{2022}\natexlab{}.
\newblock \showarticletitle{VulCNN: An Image-inspired Scalable Vulnerability Detection System}.
\newblock  (\bibinfo{year}{2022}).
\newblock


\bibitem[Yin and Ni(2024)]%
        {yin2024multitask}
\bibfield{author}{\bibinfo{person}{Xin Yin} {and} \bibinfo{person}{Chao Ni}.} \bibinfo{year}{2024}\natexlab{}.
\newblock \showarticletitle{Multitask-based Evaluation of Open-Source LLM on Software Vulnerability}.
\newblock \bibinfo{journal}{\emph{arXiv preprint arXiv:2404.02056}} (\bibinfo{year}{2024}).
\newblock


\bibitem[Zhan et~al\mbox{.}(2021)]%
        {zhan2021atvhunter}
\bibfield{author}{\bibinfo{person}{Xian Zhan}, \bibinfo{person}{Lingling Fan}, \bibinfo{person}{Sen Chen}, \bibinfo{person}{Feng We}, \bibinfo{person}{Tianming Liu}, \bibinfo{person}{Xiapu Luo}, {and} \bibinfo{person}{Yang Liu}.} \bibinfo{year}{2021}\natexlab{}.
\newblock \showarticletitle{Atvhunter: Reliable version detection of third-party libraries for vulnerability identification in android applications}. In \bibinfo{booktitle}{\emph{2021 IEEE/ACM 43rd International Conference on Software Engineering (ICSE)}}. IEEE, \bibinfo{pages}{1695--1707}.
\newblock


\bibitem[Zhang et~al\mbox{.}(2022)]%
        {zhang2022program}
\bibfield{author}{\bibinfo{person}{Quanjun Zhang}, \bibinfo{person}{Yuan Zhao}, \bibinfo{person}{Weisong Sun}, \bibinfo{person}{Chunrong Fang}, \bibinfo{person}{Ziyuan Wang}, {and} \bibinfo{person}{Lingming Zhang}.} \bibinfo{year}{2022}\natexlab{}.
\newblock \showarticletitle{Program Repair: Automated vs. Manual}.
\newblock \bibinfo{journal}{\emph{arXiv preprint arXiv:2203.05166}} (\bibinfo{year}{2022}).
\newblock


\bibitem[Zhou et~al\mbox{.}(2019)]%
        {zhou2019devign}
\bibfield{author}{\bibinfo{person}{Yaqin Zhou}, \bibinfo{person}{Shangqing Liu}, \bibinfo{person}{Jingkai Siow}, \bibinfo{person}{Xiaoning Du}, {and} \bibinfo{person}{Yang Liu}.} \bibinfo{year}{2019}\natexlab{}.
\newblock \showarticletitle{Devign: Effective vulnerability identification by learning comprehensive program semantics via graph neural networks}. In \bibinfo{booktitle}{\emph{In Proceedings of the 33rd International Conference on Neural Information Processing Systems}}. \bibinfo{pages}{10197–10207}.
\newblock


\bibitem[Zhu et~al\mbox{.}(2021)]%
        {zhu2021syntax}
\bibfield{author}{\bibinfo{person}{Qihao Zhu}, \bibinfo{person}{Zeyu Sun}, \bibinfo{person}{Yuan-an Xiao}, \bibinfo{person}{Wenjie Zhang}, \bibinfo{person}{Kang Yuan}, \bibinfo{person}{Yingfei Xiong}, {and} \bibinfo{person}{Lu Zhang}.} \bibinfo{year}{2021}\natexlab{}.
\newblock \showarticletitle{A syntax-guided edit decoder for neural program repair}. In \bibinfo{booktitle}{\emph{Proceedings of the 29th ACM Joint Meeting on European Software Engineering Conference and Symposium on the Foundations of Software Engineering}}. \bibinfo{pages}{341--353}.
\newblock


\bibitem[Ziegler et~al\mbox{.}(2019)]%
        {ziegler2019fine}
\bibfield{author}{\bibinfo{person}{Daniel~M Ziegler}, \bibinfo{person}{Nisan Stiennon}, \bibinfo{person}{Jeffrey Wu}, \bibinfo{person}{Tom~B Brown}, \bibinfo{person}{Alec Radford}, \bibinfo{person}{Dario Amodei}, \bibinfo{person}{Paul Christiano}, {and} \bibinfo{person}{Geoffrey Irving}.} \bibinfo{year}{2019}\natexlab{}.
\newblock \showarticletitle{Fine-tuning language models from human preferences}.
\newblock \bibinfo{journal}{\emph{arXiv preprint arXiv:1909.08593}} (\bibinfo{year}{2019}).
\newblock


\end{thebibliography}
